\documentclass[%
 preprint, 
 amsmath,amssymb,
 aps, physrev,
 showkeys
]{revtex4-2}

\usepackage{array}
\usepackage{graphicx}
\usepackage{dcolumn}
\usepackage{bm}
\usepackage{hyperref}
\hypersetup{hidelinks, colorlinks=true, allcolors=blue,}
\usepackage{soul, color, xcolor}
\soulregister\cite7
\soulregister\citep7
\soulregister\citet7
\soulregister\ref7

\begin{document}

\preprint{APS}

\title{\textbf{Fluid transport by a single active filament in a three-dimensional two-phase flow}}

\author{Qian Mao}
 \email{Contact author: qian.mao@univ-amu.fr}
\author{Umberto D'Ortona}
\author{Julien Favier}
\affiliation{Aix-Marseille Univ, CNRS, Centrale Med, M2P2, Marseille, France}

\date{\today}

\begin{abstract}
Micro-scale cilia play a vital role in mucociliary clearance (MCC) in the human respiratory airways. In this numerical study, we examine fluid transport driven by the active beating of a single filament immersed in a three-dimensional two-phase flow. The cilium is modeled as an elastic filament actuated by a time-varying basal angle. The two-phase flow is resolved using the Shan-Chen model in a lattice Boltzmann solver, while the two-way coupling between the filament and the fluid is treated by the immersed boundary method. Pathological conditions such as cystic fibrosis and chronic obstructive pulmonary disease are associated with drastic alterations of MCC properties, including changes in periciliary layer (PCL) thickness and the viscosity ratio between the PCL and the mucus layer (ML). Here, we systematically investigate the effects of these parameters, along with filament bending stiffness, on the beating pattern and fluid transport. Within the parameter ranges investigated, a moderate PCL thickness and viscosity ratio, together with high bending stiffness, tend to yield higher net flow rate and transport efficiency. The underlying hydrodynamic mechanisms are characterized through analyses of the beating pattern, filament dynamics, energy partition, and flow-field evolution. Two competing mechanisms are identified: the drag-elastic force balance and the viscous diffusion of momentum. Furthermore, quantitative relationships are established between flow rate and beating pattern, expressed in terms of tip amplitude and beating asymmetry.
\end{abstract}

\maketitle

\section{Introduction}

Fluid transport by cilia, which are micro-scale hair-like structures, is ubiquitous in nature \citep{barton1967analytical,faubel2016cilia,wan2019cilia,pellicciotta2020entrainment}. Among them, mucociliary clearance in respiratory airways plays a crucial role in removing foreign pollutants and inspires the design of artificial cilia for fluid pumping \citep{wanner1996mucociliary,whitesides2006origins,vilfan2010self}. In human airways, cilia are immersed in the airway surface liquid (ASL), which consists of a periciliary layer and a mucus layer. The interaction between the cilia and the ASL is inherently bidirectional. The cilia exhibit two asymmetric beating phases, which are nearly straight during the power stroke and curved during the recovery stroke. This spatial asymmetry enables continuous mucus transport. The influence of the ASL on ciliary beating cannot be neglected, especially in pathological conditions such as cystic fibrosis and chronic obstructive pulmonary disease \citep{matsui1998evidence,tarran2005normal,vanaki2020muco}. A detailed understanding of the hydrodynamic coupling between the cilia and the ASL, and its effect on fluid transport, is thus desirable for the treatment of respiratory diseases and the design of ciliated devices.

A series of experiments have been performed to investigate the complex mechanism of micro-scale mucociliary clearance (MCC) \citep{jory2019mucus,loiseau2020active,boselli2021fluid}. Nevertheless, it is challenging to carry out such experiments. As an alternative, numerical modelling has been widely used \citep{levy2014pulmonary,xu2019mathematical,vanaki2020muco,wang2022fluid,sedaghat2023nanoparticle}, which shows advantages in quantitative analysis. In some studies, the ASL was simplified to a single-phase flow \citep{jayathilake2012three,han2018spontaneous,hu2023particle}, because the cilia are almost immersed in the periciliary layer (PCL). The ability of the cilia to transport fluid was found to be dependent on various parameters. The flow rate was proportional to the beating frequency of the cilia, and inversely proportional to the viscosity of the fluid \citep{jayathilake2012three}. The flow rate was also observed to increase with ciliary density and length up to a certain point, beyond which it reached a plateau \citep{m2019three}. On the other hand, different flow patterns have been observed. With increasing ciliary density and fluid viscosity, the flow transitioned from a poorly organized (PO)/swirly (S) regime to a fully unidirectional (FU) regime characterized by unidirectional and uniform flows, which was efficient for fluid transport \citep{loiseau2020active,gsell2020hydrodynamic,mao2024hydrodynamic}. Furthermore, the phase shift between adjacent cilia can lead to metachronal waves, which significantly increases the flow rate \citep{hussong2011continuum,elgeti2013emergence,ding2014mixing,chateau2017transport,hall2020mechanics}. In these studies, the PCL is commonly simplified as a Newtonian fluid with properties similar to water, serving as a lubricant that enables cilia to beat with low viscous resistance \citep{chateau2017transport,vanaki2020muco,sedaghat2023three}. The ML exhibits non-Newtonian properties, such as yield stress, shear thinning, and viscoelasticity. The yield stress and shear thinning properties were shown to facilitate the emergence of the FU regime \citep{mao2024hydrodynamic}. The shear thinning property was found to improve the flow transport, regardless of the different Reynolds numbers \citep{wang2023generalized}. The viscoelasticity had no significant effect in the healthy state, while it significantly decreased the flow rate in the diseased state associated with cystic fibrosis \citep{choudhury2023role}.

The properties of the ASL undergo substantial alterations under diseased conditions. For instance, the viscosity of the ML increases markedly \citep{del2000effect,guo2017computational}, while the PCL becomes significantly thinner owing to airway surface dehydration \citep{matsui1997loss,livraghi2007cystic}. These pathological changes highlight the necessity of modeling the ASL as a two-phase flow in order to reliably predict fluid transport in diseased states \citep{dillon2007fluid,lukens2010using,george2019airway,modaresi2023numerical}. The PCL contains a macromolecular mesh that prevents mucus penetration into the periciliary space and leads to the formation of a distinct ML \citep{button2012periciliary}. Rather than explicitly reproducing this PCL structure in the model, this impenetrability is typically represented using numerical methods for multiphase flow. In most prior studies, two-dimensional simulations were conducted. These works consistently showed that increasing ML viscosity reduces the flow rate relative to the healthy state \citep{lee2011muco,kurbatova2015model,guo2017computational}. A moderate ratio of PCL thickness to ciliary length was found to be favorable for fluid transport \citep{sedaghat2016numerical,guo2017computational,sedaghat2021nonlinear}. Specifically, maximum flow rates were reported when the cilia penetrated slightly into the ML \citep{jayathilake2015numerical}. Moreover, the mucus velocity was largely insensitive to ML thickness once the ML exceeded approximately half the PCL thickness \citep{smith2007discrete,sedaghat2016numerical}. These findings have also been corroborated by three-dimensional simulations \citep{chatelin2013hybrid,chateau2017transport,m2019three,sedaghat2022nonlinear}. Nevertheless, two-dimensional models generally overestimate the flow rate \citep{vanaki2020muco}. Furthermore, most existing simulations employed prescribed ciliary beating patterns \citep{chatelin2016parametric,chateau2017transport,chateau2019antiplectic,quek2018three,sedaghat2023three}, implying a one-way coupling between the cilia and the two-phase flow that neglected fluid feedback on ciliary deformation. In pathological ASL, characterized by elevated ML viscosity and reduced PCL thickness, the fluid feedback can induce ciliary deformation, thereby impairing MCC \citep{guo2017computational,vanaki2020muco}. Only a few studies \citep{mitran2007metachronal,dillon2007fluid,dillon2000integrative,lukens2010using,stein2019coarse,yang2008integrative} have incorporated two-way coupling in modeling the cilia-mucus system, primarily due to its numerical complexity. Most of these works employed two-dimensional models \citep{dillon2007fluid,dillon2000integrative,lukens2010using,yang2008integrative}. In particular, \citet{mitran2007metachronal} developed a complex two-way coupling model to investigate ciliary metachronal waves in three-dimensional two-phase flow, in which the cilium was modeled with a 9 + 2 internal microtubule structure capable of large deflections. More recently, \citet{stein2019coarse} proposed a simplified two-way coupling framework based on the Brinkman-Elastica model derived from local slender-body theory, allowing efficient simulations of dense ciliary arrays. However, these studies kept the PCL thickness and ML viscosity fixed, leaving their roles in regulating fluid transport and ciliary dynamics within a three-dimensional two-way coupled system unclear. The present study focuses first on these properties of the ASL, even though it is well known that the ML also exhibits a complex non-Newtonian behavior, which is difficult to handle both experimentally and numerically \citep{lafforgue2017thermo}.

The objective of the present study is to numerically investigate the fluid transport by an active filament in a three-dimensional two-phase flow. The two-phase flow is simulated using the Shan-Chen model in a lattice Boltzmann solver. The dynamics of the beating cilium is modeled by an elastic filament driven by a time-varying basal angle ($\theta(t)$). The interaction between the two-phase flow and the filament is treated by the immersed boundary (IB) method. The beating pattern and the filament-driven fluid transport are first analysed in a representative case. The effects of the PCL thickness ($L_{\mathrm{PCL}}$), the viscosity ratio ($r_{\nu}$) between the ML and the PCL, and the bending stiffness ratio ($r_B$ that will be defined soon) on the fluid transport are explored. The corresponding hydrodynamic mechanism is characterized through an analysis of the filament beating pattern, the variations in the hydrodynamic force and energy components, and the evolution of the flow field. The relationship between the flow rate and the beating pattern is elucidated.

\section{Computational model}
\label{sec:computational_model}

\begin{figure}
\centerline{\includegraphics[width=0.75\linewidth]{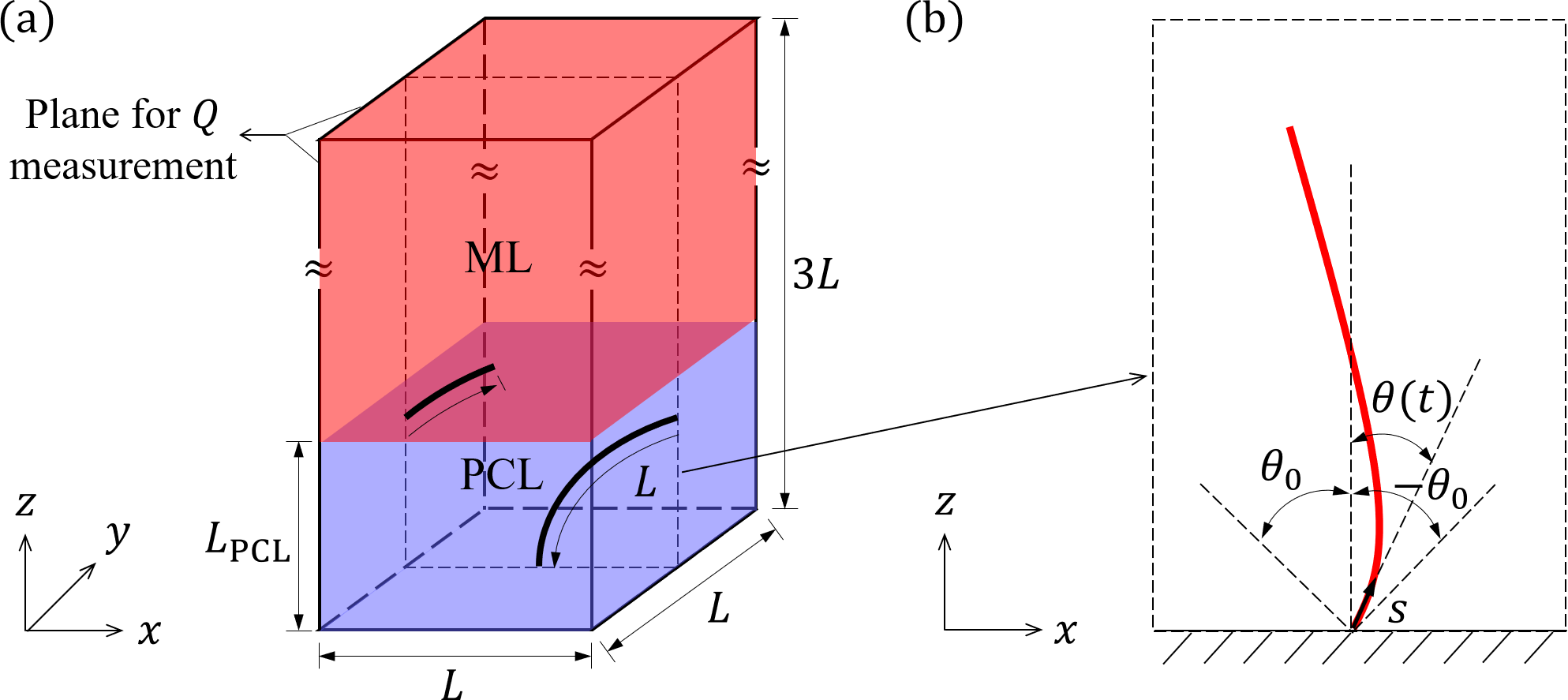}}
\caption{(a) Schematic of a filament in the computational domain (not in scale), where the filament beats in the plane $y=0.5L$. (b) Schematic of a beating filament actuated by its basal angle ($\theta(t)$).}
\label{fig_schem}
\end{figure}

A schematic of a filament immersed in a three-dimensional two-phase flow is shown in Fig. \ref{fig_schem}(a), where the periciliary layer (PCL) is colored in blue and the mucus layer (ML) is colored in red. The length of the filament is $L$. The computational domain is $L$ in length and width, and $3L$ in height. The thickness of the PCL is $L_{\mathrm{PCL}}$. A two-way coupling between the fluid motion and the filament motion is realized, i.e. the beating of the filament drives the fluid flow, and the fluid flow in turn affects the deformation of the filament. The fluid motion is defined on a fixed Cartesian grid. A no-slip boundary condition is applied at the bottom boundary of the domain. A free slip boundary condition is specified at the top boundary. A periodic boundary condition is imposed to the other four boundaries. The initial condition is $\boldsymbol{u}=0$. The filament is forced to move in the plane $y=0.5L$ and its base is fixed at ($0.5L$, $0.5L$, 0). Figure \ref{fig_schem}(b) shows a schematic of a beating filament. The motion of the filament is defined on a moving Lagrangian grid under a curvilinear coordinate system ($s$). The basal angle ($\theta(t)$) of the filament is varied sinusoidally to realize a periodic beating.

The fluid motion is governed by the incompressible Navier-Stokes (NS) equations, which is solved by the lattice Boltzmann method (LBM) \citep{Kruger2017lattice}. The lattice velocity is $c=\Delta x/\Delta t$ ($\Delta x=1$, $\Delta t=1$). The lattice Boltzmann equation normalized by $c$ and $\Delta t$ is written as
\begin{equation}
\displaystyle
    f^{\sigma}_l(\boldsymbol{x}+\boldsymbol{e}_l,t+1) -f^{\sigma}_l(\boldsymbol{x},t)=\Gamma^{\sigma}_{l}(\boldsymbol{x},t) + S^{\sigma \prime}_l(\boldsymbol{x},t),
    \label{discret_lbm_force}
\end{equation}
where $f^{\sigma}_l(\boldsymbol{x},t)$ is the distribution function of the $\sigma$th fluid component ($\sigma =$ 1 and 2) and $\boldsymbol{e}_l$ ($l=0$, ..., $Q-1$) are the velocity vectors, where $Q$ is the number of discrete velocities. The $D3Q19$ scheme is employed in the present study \citep{Kruger2017lattice}. $\Gamma^{\sigma}_{l}$ is the collision operator and the two-relaxation-time (TRT) collision model is used:
\begin{equation}
    \Gamma^{\sigma}_l = -\frac{1}{\tau^+_{\sigma}}(f^{\sigma +}_l-f^{\sigma (eq)+}_l)-\frac{1}{\tau^-_{\sigma}} (f^{\sigma -}_l-f^{\sigma (eq)-}_l),
    \label{TRT_coll_step}
\end{equation}
where the superscripts $+$ and $-$ represent the symmetric and anti-symmetric parts of the quantities. $\tau_{\sigma}$ is the relaxation time and $\tau^+_{\sigma}$ determines the kinematic fluid viscosity ($\nu_{\sigma}=c_s^2(\tau^+_{\sigma} - 1/2)$). Here, $c_s=\Delta x/(\sqrt{3}\Delta t)$ is the lattice sound speed. $\tau^-_{\sigma}$ is a numerical parameter and controlled by $\Lambda_\tau=(\tau^+_{\sigma} -1/2)(\tau^-_{\sigma} -1/2)=1/4$ \citep{ginzburg2010optimal}.

The equilibrium distribution function $f_l^{\sigma (eq)}$ is calculated by
\begin{equation}
\begin{array}{c}
\displaystyle
f_l^{\sigma (eq)}(\boldsymbol{x},t) = w_l\rho_{\sigma} \Big[1+\frac{\boldsymbol{e}_l \cdot \boldsymbol{u}^{eq}_{\sigma}}{c_s^2}+\frac{(\boldsymbol{e}_l \cdot \boldsymbol{u}^{eq}_{\sigma})^2}{2c_s^4}-\frac{\boldsymbol{u}^{eq}_{\sigma} \cdot \boldsymbol{u}^{eq}_{\sigma}}{2c_s^2}\Big],
\end{array}{}
\label{pop_equilibre}
\end{equation}
where $w_l$ are the lattice weights \citep{Kruger2017lattice}. The equilibrium velocity ($\boldsymbol{u}^{eq}$) is the same for each component ($\boldsymbol{u}^{eq}_{\sigma}=\boldsymbol{u}^{eq}$):
\begin{equation}
\displaystyle
    \boldsymbol{u}^{eq}= \frac{\displaystyle \sum_{\sigma} {\rho}_{\sigma} \boldsymbol{u}_{\sigma} / \tau^+_{\sigma}}{\displaystyle \sum_{\sigma} {\rho}_{\sigma} / \tau^+_{\sigma}},
    \label{eq_velo}
\end{equation}
where $\boldsymbol{u}_{\sigma}$ is the flow velocity of the $\sigma$th fluid component. The symmetric and anti-symmetric parts of $f_l^{\sigma}$ and $f_l^{\sigma (eq)}$ are defined as
\begin{equation}
\left\{   
    \begin{array}{cc}
         f_l^{\sigma +}=\dfrac{f_l^{\sigma}+f_{\bar{l}}^{\sigma}}{2}, & f_l^{\sigma -}=\dfrac{f_l^{\sigma}-f_{\bar{l}}^{\sigma}}{2},\\
         \\
         f_l^{\sigma (eq)+}=\dfrac{f_l^{\sigma (eq)}+f_{\bar{l}}^{\sigma (eq)}}{2}, & f_l^{\sigma (eq)-}=\dfrac{f_l^{\sigma (eq)}-f_{\bar{l}}^{\sigma (eq)}}{2},\\ 
   \end{array}{}
\right.
\label{trt_1}
\end{equation}
where the subscript $\bar l$ is defined such that $\boldsymbol{c_{\bar l}}=-\boldsymbol{c_l}$. The external body force term ($S^{\sigma \prime}_l$) is expressed as
\begin{equation}
\displaystyle
    S^{\sigma \prime}_l=\Big(1-\frac{1}{2\tau_{\sigma}^+}\Big) S^{\sigma +}_l + \Big(1-\frac{1}{2\tau_{\sigma}^-}\Big) S^{\sigma -}_l,
    \label{force_guo_trt}
\end{equation}
where $S^{\sigma +}_l=(S_l^{\sigma}+S_{\bar{l}}^{\sigma})/2$ and $S^{\sigma -}_l=(S_l^{\sigma}-S_{\bar{l}}^{\sigma})/2$ are the symmetric and anti-symmetric parts of $S_l^{\sigma}$, which is defined as
\begin{equation}
\displaystyle
    S_l^{\sigma}=w_l\Big[ \frac{\boldsymbol{e}_l-\boldsymbol{u}_{\sigma}^{eq}}{c_s^2}+ \frac{(\boldsymbol{e}_l \cdot \boldsymbol{u}_{\sigma}^{eq}) \boldsymbol{e}_l}{c_s^4}\Big] \boldsymbol{f}^{\mathrm{EXT}}_{\sigma},
    \label{force_guo}
\end{equation}
where $\boldsymbol{f}^{\mathrm{EXT}}_{\sigma}$ is the sum of the Shan-Chen-type fluid-fluid cohesion force ($\boldsymbol{f}^{\mathrm{SC}}_{\sigma}$) for modelling two-phase flows and the IB-related force ($\boldsymbol{f}^{\mathrm{IB}}_{\sigma}$) exerted by the filament on the surrounding fluid. $\boldsymbol{f}^{\mathrm{SC}}_{\sigma}$ is calculated by \citep{li2016immersed}
\begin{equation}
         \boldsymbol{f}^{\mathrm{SC}}_{\sigma}(\boldsymbol{x},t)=-G_{coh}\rho_{\sigma}(\boldsymbol{x},t)\displaystyle \sum_l w_l \rho_{\sigma ^{\prime}}(\boldsymbol{x}+\boldsymbol{e}_l, t) \boldsymbol{e}_l,
     \label{shan-chen}     
\end{equation}
where $G_{coh}$ is the parameter that controls the strength of $\boldsymbol{f}^{\mathrm{SC}}_{\sigma}$ \citep{huang2007proposed}, and $\sigma ^{\prime}$ represents another fluid component different from $\sigma$. The PCL and ML are treated as immiscible phases, with $G_{coh}=1.8$ chosen to achieve stable phase separation while maintaining the surface tension as low as possible \citep{huang2007proposed}. Surface tension is implicitly represented and controlled by $G_{coh}$. This model allows the interface between the PCL and ML to evolve. However, in the present study, the movement of the interface remains minor under the influence of a single filament.

In this study, we simulate a two-phase flow with various viscosity ratios. A system relaxation time ($\tau _{sys}$) is defined \citep{zhao2021modified} to calculate the relaxation time and the viscosity depending on the fluid nature at each node, ensuring a good numerical stability:
\begin{equation}
\displaystyle
    \tau _{sys} = \frac{\displaystyle \sum_{\sigma} \rho_{\sigma} \nu_{\sigma}}{\displaystyle \sum_{\sigma} \rho_{\sigma} c^2_s} + \frac{1}{2},
\label{eq_tausys}
\end{equation}
where $\rho_{\sigma}$ is the fluid density of the $\sigma$th fluid component. Then, $\tau_{\sigma}^+$ in Eqs. (\ref{TRT_coll_step}), (\ref{eq_velo}) and (\ref{force_guo_trt}) is replaced by $\tau _{sys}$ for each component. $\tau_{\sigma}^-$ is recalculated according to $\Lambda_\tau$.

The fluid density ($\rho_{\sigma}$) and velocity ($\boldsymbol{u}_{\sigma}$) of the $\sigma$th fluid component are given by
\begin{equation}
    \displaystyle
         \rho _{\sigma}=\displaystyle \sum_l f_l^{\sigma},
     \label{density}
\end{equation} 

\begin{equation}
    \displaystyle
         \rho _{\sigma}\boldsymbol{u}_{\sigma}=\displaystyle \sum_l f_l^{\sigma} \boldsymbol{e}_l + \dfrac{1}{2} \boldsymbol{f}_{\sigma}^{\mathrm{EXT}}.
     \label{momentum}
\end{equation} 

Note that the real flow velocity ($\boldsymbol{u}$) is equal to the equilibrium velocity $\boldsymbol{u}^{eq}$. The fluid density $\rho _\mathrm{f}$ is equal to $\sum \rho_{\sigma}$. The densities of the PCL and the ML are identical. The current method of simulating two-phase flows has been carefully validated in our previous studies \citep{li2016immersed,chateau2017transport,chateau2019antiplectic}.

The filament motion is governed by the following equation and the inextensibility condition \citep{huang2007simulation,favier2014lattice,mao2023drag}:
\begin{equation}
\displaystyle
    \rho_{\mathrm{s}} \frac{\partial^2 \boldsymbol{X}}{\partial t^{2}} = \frac{\partial}{\partial s}\Big(T\frac{\partial \boldsymbol{X}}{\partial s}\Big) - \frac{\partial^2}{\partial s^{2}}\Big(B\frac{\partial^2 \boldsymbol{X}}{\partial s^{2}}\Big) - \boldsymbol{F}_s^{\mathrm{IB}},
    \label{eq_solidmotion}
\end{equation}

\begin{equation}
\displaystyle
    \frac{\partial \boldsymbol{X}}{\partial s} \cdot \frac{\partial \boldsymbol{X}}{\partial s} = 1,
    \label{eq_inexten}
\end{equation}
where $t$ is time, $s$ ranges from 0 to $L$, $\boldsymbol{X}=(X(s,t), 0.5L, Z(s,t))$ is the position of the filament, $\boldsymbol{F}_s^{\mathrm{IB}}$ is the IB-related force per unit length exerted by the fluid, $\rho_{\mathrm{s}}$ is the linear density of the filament (solid), $T$ is the tension coefficient along the filament's axis and is determined by the constraint of inextensibility \citep{huang2007simulation}, and $B$ is the bending stiffness. For further information, please refer to the preceding works \citep{huang2007simulation,favier2014lattice}.

The periodic beating of the filament is modeled by varying the basal angle $\theta(t)$ sinusoidally:
\begin{equation}
    \theta(t)=\begin{cases}\theta_0 \cos(\pi f_\mathrm{b} t/r_{T_\mathrm{bp}}) & \text{ Power stroke } \\
    \\-\theta_0 \cos(\pi f_\mathrm{b} (t-r_{T_\mathrm{bp}} T_\mathrm{b})/(1-r_{T_\mathrm{bp}})) & \text{ Recovery stroke } \end{cases}
\label{eq_Theta}
\end{equation}
where $\theta_0$ is the angular amplitude, $f_{\mathrm{b}}$ is the beating frequency, $r_{T_\mathrm{bp}} = T_\mathrm{bp}/T_\mathrm{b}$ is the ratio between the duration of the power stroke ($T_\mathrm{bp}$) and the duration of one beating period ($T_\mathrm{b}$). $T_\mathrm{br}$ is the duration of the recovery stroke.

Again, the cilium exhibits two asymmetric beating phases, i.e. nearly straight during the power stroke and curved during the recovery stroke. A simple and effective way to reproduce its basic beating pattern is to prescribe a time-varying bending stiffness ($B(t)$):
\begin{equation}
   B(t)=\begin{cases}B_{\mathrm{max}} & \text{ Power stroke } \\
    \\B_{\mathrm{min}}+(B_{\mathrm{max}}-B_{\mathrm{min}})\left(\frac{t-T_\mathrm{bp}}{T_\mathrm{br}}\right)^n & \text{ Recovery stroke } \end{cases}
\label{eq_B}
\end{equation}
where $B_{\mathrm{max}}$ and $B_{\mathrm{min}}$ denote the maximum and minimum bending stiffness, respectively, and $n$ is the power index. The filament is stiff during the power stroke ($B(t)=B_{\mathrm{max}}$). Upon entering the recovery stroke, it softens ($B(t)$ decreases to $B_{\mathrm{min}}$), after which its stiffness gradually recovers to $B_{\mathrm{max}}$. Here, $B_{\mathrm{min}}^*=40$ and $n=12$ are empirical parameters determined by trial and error, chosen to qualitatively reproduce the ciliary beating pattern. Varying $B_{\mathrm{min}}^*$ between 20$-$60 or $n$ between 10$-$14 results in only minor changes in the beating amplitude. The resulting beating pattern is generally consistent with experimental observations \citep{gheber1997extraction, chioccioli2019quantitative}. It should be noted that this simplified model is designed only to capture the essential feature of asymmetric ciliary beating without attempting to match the exact motion. The underlying mechanism and detailed modeling of cilium motion remain active topics of ongoing research.

A clamped boundary condition is applied to the base of the filament, while the other parts of the filament are free to move. The clamped boundary conditions are given by
\begin{equation}
    \begin{cases}\frac{\partial X(0,t)}{\partial s} = \cos[\theta(t) + \pi/2]\\
    \\\frac{\partial Z(0,t)}{\partial s} = \sin[\theta(t) + \pi/2]\end{cases}.
\label{eq_clamped}
\end{equation}

The interaction between the fluid and filament motions is treated using the IB method. The IB-related force exerted by each fluid component ($\boldsymbol{F}_{\sigma,s}^{\mathrm{IB}'}$) is obtained from \citep{li2016immersed}
\begin{equation}
    \mathcal{I}[\rho_{\sigma}]_{s}\boldsymbol{u}_{\sigma,s}=\mathcal{I}\left[\sum_l f_{l}^{\sigma}\boldsymbol{e}_{l}\right]_{s}+\mathcal{I}\left[\frac{1}{2}\boldsymbol{F}_{\sigma}^{\mathrm{SC}}\right]_{s}+\frac{1}{2}\boldsymbol{F}_{\sigma,s}^{\mathrm{IB}'},
    \label{eq_FIB}
\end{equation}
where $\boldsymbol{u}_{\sigma,s}$ is the flow velocity of the $\sigma$th fluid component at the $s$th Lagrangian point, which is equal to the velocity of the solid boundary to enforce the no-slip boundary condition along the IB. $\mathcal{I}[\bullet]_{s}$ represents the interpolation operator:
\begin{equation}
    \phi(\boldsymbol{x}_s,t)=\mathcal{I}[\phi(\boldsymbol{x},t)]_s=\int\phi(\boldsymbol{x},t)\delta(\boldsymbol{x}-\boldsymbol{x}_s) d\boldsymbol{x},
\end{equation}
where $\delta$ is the Dirac delta function \citep{roma1999adaptive,li2016immersed}:

\begin{equation}
\delta(\boldsymbol{x}-\boldsymbol{x}_{s})=\frac{1}{\Delta x^3}\widetilde{\delta}\left(\frac{{x}-x_{s}}{\Delta x}\right)\widetilde{\delta}\left(\frac{{y}-y_{s}}{\Delta x}\right)\widetilde{\delta}\left(\frac{{z}-z_{s}}{\Delta x}\right),
\end{equation}

\begin{equation}
\widetilde{\delta}(r)=\begin{cases}\frac{1}{3}(1+\sqrt{1-3r^2}),&|r|\leqslant\frac{1}{2}\\\frac{1}{6}(5-3|r|-\sqrt{-2+6|r|-3r^2}),&\frac{1}{2}<|r|<\frac{3}{2},\\0,&|r|\geqslant\frac{3}{2}\end{cases}
\end{equation}

The force is then corrected by a coefficient ($\boldsymbol{F}_{\sigma,s}^{\mathrm{IB}} = W_{\sigma} \boldsymbol{F}_{\sigma,s}^{\mathrm{IB}'}$) to accurately impose the no-slip condition at the filament and to mitigate spurious velocity oscillations that grow with larger relaxation time (viscosity) \citep{gsell2019explicit}. The coefficient $W_{\sigma}$ is defined as
\begin{equation}
    W_{\sigma}=\frac{\lambda_{\sigma}}{1+\kappa(\lambda_{\sigma}-1)},
    \label{eq_WIB}
\end{equation}
where $\lambda_{\sigma}$ is defined as $\lambda_{\sigma}=2\tau_{\sigma}^+ -1$, and $\kappa$ is 0.5 for the current delta function. The total IB-related force exerted on the filament is $\boldsymbol{F}_s^{\mathrm{IB}} = \sum \boldsymbol{F}_{\sigma,s}^{\mathrm{IB}}$. After that, $\boldsymbol{f}_{\sigma}^{\mathrm{IB}}$ is obtained by spreading $\boldsymbol{F}_{\sigma,s}^{\mathrm{IB}}$ from the Lagrangian points to the $j$th Eulerian node \citep{li2016immersed}:
\begin{equation}
    \boldsymbol{f}_{\sigma}^{\mathrm{IB}}(\boldsymbol{x}_{j},t)=\mathcal{S}[\boldsymbol{F}_{\sigma}^{\mathrm{IB}}(\boldsymbol{x}_{s},t)]_j=\int \boldsymbol{F}_{\sigma}^{\mathrm{IB}}(\boldsymbol{x}_{s},t)\delta(\boldsymbol{x}_{j}-\boldsymbol{x}_{s}) ds.
    \label{eq_fIB_eu}
\end{equation}

Details on the current IB method and its validation can be found in our previous works \citep{favier2014lattice,li2016immersed,mao2023snap,mao2023drag2}.

The computational procedure of the present numerical framework can be summarized as follows. (i) At the $n$th time step, the fluid velocity ($\boldsymbol{u}_{\sigma}^n$) and density ($\rho_{\sigma}^n$) fields and the position of the filament $\boldsymbol{X}^n$ are known. Calculate IB force $\boldsymbol{F}_{\sigma}^{\mathrm{IB}}$ by Eqs. (\ref{eq_FIB}) and (\ref{eq_WIB}), and spread it to the Eulerian node by Eq. (\ref{eq_fIB_eu}). Calculate the Shan-Chen-type fluid-fluid cohesion force $\boldsymbol{f}^{\mathrm{SC}}_{\sigma}$ by Eq. (\ref{shan-chen}). (ii) Solve Eq. (\ref{discret_lbm_force}) to obtain the distribution function $f^{\sigma,n+1}_l$ at new time step. Obtain fluid velocity ($\boldsymbol{u}_{\sigma}^{n+1}$) and density ($\rho_{\sigma}^{n+1}$) fields at new time step by Eqs. (\ref{density}) and (\ref{momentum}). (iii) Solve Eqs. (\ref{eq_solidmotion}) and (\ref{eq_inexten}) to obtain the filament position ($\boldsymbol{X}^{n+1}$) at new time step. This ends one time step marching.

\begin{figure}
\centerline{\includegraphics[width=0.9\linewidth]{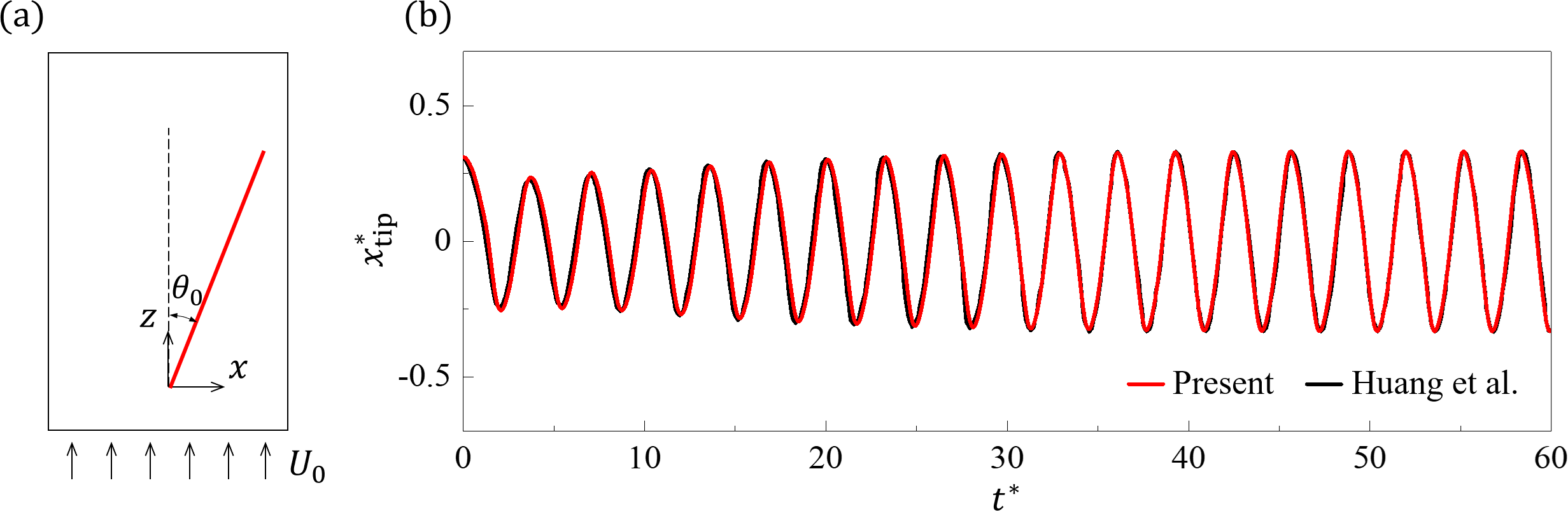}}
\caption{(a) Schematic of a filament in an uniform flow (not in scale). (b) Time histories of the $x$-position of the filament tip ($x_{\mathrm{tip}}^*$).}
\label{fig_benchmark}
\end{figure}

To preliminary validate the present framework, we simulate a flexible filament subjected to gravity and immersed in a two-dimensional uniform single-phase flow, as shown in Fig. \ref{fig_benchmark}(a). The two-dimensional flow is mimicked by using a single grid in the $y$-direction, with periodic boundary conditions applied at the front and back boundaries. A velocity inlet is imposed at the bottom boundary, while a Neumann boundary condition is applied at the top. Slip boundary conditions are specified at the left and right boundaries. The computational domain extends over $-4L \leq x \leq 4L$ and $-2L \leq z \leq 6L$. The filament base is fixed at $(0,0)$, with an initial angle of $\theta_0=0.1\pi$. The bending stiffness and mass ratio are set to $B^*=B/(\rho_\mathrm{s}U_\mathrm{0}^2L^2)=0.001$ and $m^*=\rho_\mathrm{s}/(\rho_\mathrm{f}L^2)=1.5$, respectively. The Reynolds number is $Re=U_0L/\nu=200$. To account for gravity, a body-force term ($\rho_\mathrm{s}\boldsymbol{g}$) is added to the right-hand side of Eq. (\ref{eq_solidmotion}). The gravity is applied in the direction of the inflow, and its magnitude is controlled by the Froude number $Fr=gL/U_\mathrm{0}^2=0.5$. The details of this benchmark configuration follow \citet{huang2007simulation}. Figure \ref{fig_benchmark}(b) compares the time history of the filament tip $x$-position, showing good agreement with the reference result, thereby validating the present framework.

\begin{table}[b]
\caption{Definitions and selected values of primary dimensionless parameters. Symbol '$-$' indicates that the corresponding parameter is updated during the simulation.}
\begin{ruledtabular}
\begin{tabular}{lcc}
\textrm{Dimensionless parameter}&
\textrm{Definition}&
\textrm{Value}\\
\colrule
Filament length & $L^*=L/L$ & 1 \\
Eulerian coordinate & $\boldsymbol{x}^*=\boldsymbol{x}/L$ & $-$ \\
Lagrangian coordinate & $s^*=s/L$ & $-$ \\
Filament position & $\boldsymbol{X}^*=\boldsymbol{X}/L$ & $-$ \\
Domain length & $L_{\mathrm{d}}^*=L_{\mathrm{d}}/L$ & 1 \\
Domain width & $W_{\mathrm{d}}^*=W_{\mathrm{d}}/L$ & 1 \\
Domain height & $H_{\mathrm{d}}^*=H_{\mathrm{d}}/L$ & 3 \\
PCL thickness & $L_{\mathrm{PCL}}^*=L_{\mathrm{PCL}}/L$ & 0.6$-$1.1 \\
Beating period & $T_{\mathrm{b}}^*=T_{\mathrm{b}}U_{\mathrm{r}}/L$ & 2 \\
Beating frequency & $f_{\mathrm{b}}^*=1/T_{\mathrm{b}}^*$ & 0.5 \\
Power stroke proportion & $r_{T_\mathrm{bp}}$ & $1/3$ \\
Angular amplitude & $\theta_0$ & $\pi /3$ \\
Time & $t^*=t/T_{\mathrm{b}}$ & $-$ \\
Mass ratio & $m^*=\rho_{\mathrm{s}}/\rho_{\mathrm{f}}L^2$ & $8.72 \times 10^{-3}$ \\
Viscosity ratio & $r_{\nu}=\nu_{\mathrm{ML}}/\nu_{\mathrm{PCL}}$ & 1$-$50 \\
Bending stiffness ratio & $r_B=B_{\mathrm{max}}/B_{\mathrm{min}}$ & 10$-$70 \\
Bending stiffness & $B^*=B/\rho_{\mathrm{s}} U_{\mathrm{r}}^2 L^2$ & $-$ \\
Minimum bending stiffness & $B_{\mathrm{min}}^*=B_{\mathrm{min}}/\rho_{\mathrm{s}} U_{\mathrm{r}}^2 L^2$ & 40 \\
Power index & $n$ & 12 \\
Flow velocity & $\boldsymbol{u}^*=\boldsymbol{u}/U_{\mathrm{r}}$ & $-$ \\
Reynolds number & $Re=U_{\mathrm{r}}L/\nu_{\mathrm{PCL}}$ & 0.1 \\
IB-related force & $\boldsymbol{F}_s^{\mathrm{IB}*}=\boldsymbol{F}_s^{\mathrm{IB}}L/\rho_{\mathrm{s}}U_{\mathrm{r}}^2$ & $-$ \\
\end{tabular}
\end{ruledtabular}
\label{table01}
\end{table}

For ease of reference, table \ref{table01} summarizes the definitions and selected values of the primary dimensionless parameters. The Reynolds number $Re=U_{\mathrm{r}} L / \nu_{\mathrm{PCL}}$ is fixed at 0.1 to prevent inertial effects \citep{loiseau2020active}, where $U_{\mathrm{r}}$ is the reference velocity and $\nu_{\mathrm{PCL}}$ is the kinematic viscosity of the PCL. In lab-on-a-chip devices, $Re$ typically lies in the range $O(0.01)-O(1)$ \citep{shields2010biomimetic,hanasoge2018microfluidic,milana2020metachronal}. Experimental measurements report a PCL thickness of 6$-$7 $\mathrm{\mu m}$ \citep{matsui1998evidence} and a cilium length of approximately 8 $\mathrm{\mu m}$ \citep{chatelin2016parametric}, corresponding to a dimensionless PCL thickness of $L_{\mathrm{PCL}}^* = L_{\mathrm{PCL}}/L \approx 0.75-$0.875. In the present study, $L_{\mathrm{PCL}}^*$ is varied between 0.6 and 1.1 to account for pathological conditions. The cilium beating frequency is 10 Hz \citep{chateau2017transport}, and the typical mucus velocity is approximately $1.783\times 10^{-4}$ m/s \citep{morgan2004scintigraphic}, yielding a dimensionless beating period of $T_{\mathrm{b}}^* \approx 2$ and a dimensionless beating frequency of $f_{\mathrm{b}}^* \approx 0.5$. The ratio between the duration of the power stroke and the beating period is fixed at $r_{T_\mathrm{bp}}=1/3$ \citep{widdicombe1995regulation,chateau2017transport}, and the angular amplitude is set to $\theta_0=\pi/3$ \citep{sleigh1988propulsion,chateau2017transport}. The densities of the cilium and mucus are 1110 and 1000 kg/m$^3$, respectively \citep{cui2022three}. The aspect ratio of the cilium (diameter-to-length) is about 0.03$-$0.05 in experiments \citep{chateau2017transport}; here, a value of 0.1 is used to satisfy the slender body condition, yielding a mass ratio of $m^* \approx 8.72 \times 10^{-3}$. Note that $\rho_{\mathrm{s}}$ is the linear density of the filament. The viscosity of healthy mucus ranges from $5\times 10^{-3}$ to $5\times 10^{-2} $ Pa$\cdot$s \citep{loiseau2020active}, whereas the PCL has a viscosity of approximately $1\times 10^{-3} $ Pa$\cdot$s \citep{chatelin2016parametric}, resulting in a viscosity ratio ($r_{\nu}=\nu_{\mathrm{ML}}/\nu_{\mathrm{PCL}}$) of 5$-$50. \citet{chatelin2016parametric} reported that the fluid transport is maximized when $r_{\nu}$ lies between 10 and 20. In the present study, $r_{\nu}$ is varied from 1 to 50. The bending stiffness ratio ($r_B$) ranges from 10 to 70.

The filament dynamics and fluid transport in the present study are quantified by several dimensionless quantities. The dimensionless force exerted by the surrounding fluid on the filament is calculated as
\begin{equation}
    \boldsymbol{F}^{\mathrm{IB} *}= \int - \boldsymbol{F}_s^{\mathrm{IB} *}ds^*.
\end{equation}

The time-averaged dimensionless flow rate is defined as
\begin{equation}
    \bar{Q}^* = \frac{1}{T^*} \int Q^* d t^*,
    \label{eq_aveflowrate}
\end{equation}
where $Q^*= \iint u^* dy^*dz^*$ is the instantaneous dimensionless flow rate across the plane $x^*=0$ as shown in Fig. \ref{fig_schem}, where $u^*=u/U_{\mathrm{r}}$ is the dimensionless flow velocity. $\bar{Q}^*_{\mathrm{ML}}$ is the time-averaged dimensionless flow rate of the ML. The time-averaged dimensionless kinetic energy ($\bar{E}_{\mathrm{ks}}^*$) of the filament is defined as
\begin{equation}
    \bar{E}_{\mathrm{ks}}^* = \frac{1}{T^*} \int E_{\mathrm{ks}}^* d t^*,
    \label{eq_aveek}
\end{equation}
where $E_{\mathrm{ks}}^* = \int 0.5 \rho_{\mathrm{s}}^{*} \boldsymbol{U}_s^{*2} ds^*$ is the instantaneous dimensionless kinetic energy, where $\rho_{\mathrm{s}}^{*}$ is the dimensionless density and $\boldsymbol{U}_s^*=d\boldsymbol{X}^*/dt^*$ is the dimensionless velocity of the filament at the $s$th Lagrangian point. The time-averaged dimensionless elastic strain energy ($\bar{E}_{\mathrm{es}}^*$) of the filament is defined as
\begin{equation}
    \bar{E}_{\mathrm{es}}^* = \frac{1}{T^*} \int E_{\mathrm{es}}^* d t^*,
    \label{eq_avees}
\end{equation}
where $E_{\mathrm{es}}^*$ is given by
\begin{equation}
    E_{\mathrm{es}}^* = \int \frac{1}{2} B^* \Big(\frac{\partial^2 \boldsymbol{X}^*}{\partial s^{*2}}\Big)^2 ds^*.
    \label{eq_esfilament}
\end{equation}

The time-averaged dimensionless input power ($\bar{P}_{\mathrm{in}}^*$) of the filament is defined as
\begin{equation}
    \bar{P}_{\mathrm{in}}^* = \frac{1}{T^*} \int P_{\mathrm{in}}^* d t^*,
    \label{eq_avepin}
\end{equation}
where $P_{\mathrm{in}}^*$ is defined as the power required to produce the filament beating:
\begin{equation}
    P_{\mathrm{in}}^* = \int \boldsymbol{F}^{\mathrm{IB} *} \cdot \boldsymbol{U}_s^* ds^*.
    \label{eq_pinfilament}
\end{equation}

Part of the input power is converted into the kinetic and elastic strain energy of the filament, while the remainder is dissipated into the surrounding fluid. It should be noted that the dissipated power is not entirely effective in driving forward transport, since the filament also generates a backward flow during the recovery stroke. To quantify the effective contribution, we define a dimensionless time-averaged kinetic energy of the fluid ($\bar{E}_{\mathrm{kf}}^*$), which characterizes the forward-directed kinetic energy of the fluid:
\begin{equation}
    \bar{E}_{\mathrm{kf}}^* = \frac{\bar{u}^{*}}{|\bar{u}^{*}|} \int 0.5 \rho_{\mathrm{f}}^{*} \bar{u}^{*2} dx^{*3},
    \label{eq_aveekf}
\end{equation}
where $\rho_{\mathrm{f}}^{*}$ is the dimensionless density and $\bar{u}^*$ denotes the time-averaged streamwise flow velocity. Based on this, the transport efficiency ($\eta$) is defined as
\begin{equation}
    \eta = \frac{\bar{E}_{\mathrm{kf}}^*}{\bar{P}_{\mathrm{in}}^*}.
    \label{eq_energyeffi}
\end{equation}

\begin{figure}
\centerline{\includegraphics[width=\linewidth]{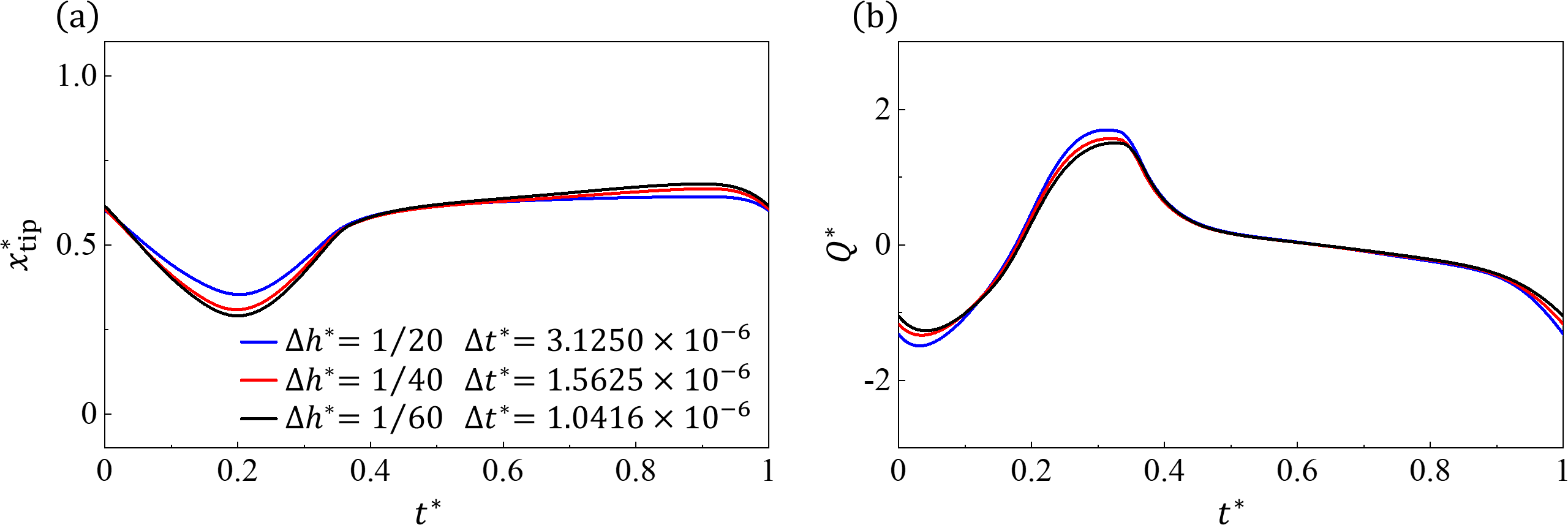}}
\caption{Time histories of (a) the $x$-position of the filament tip ($x_{\mathrm{tip}}^*$) and (b) the flow rate ($Q^*$) in one beating period for different grid sizes $ \Delta h^*$ and time steps $\Delta t^*$ (PCL thickness $L_\mathrm{PCL}^*=0.9$, viscosity ratio $r_\nu = 50$, bending stiffness ratio $r_B = 10$).}
\label{fig_gridtest}
\end{figure}

\begin{table}[b]
\caption{Tip amplitude ($\Delta x_{\mathrm{tip}}^*$), time-averaged flow rate ($\bar{Q}^*$), and relative errors ($err$) to $\Delta h^*=1/60$ for different gird sizes ($\Delta h^*$).}
\begin{ruledtabular}
\begin{tabular}{lcccc}
\textrm{$\Delta h^*$}&
\textrm{$\Delta x_{\mathrm{tip}}^*$}&
\textrm{$err$ ($\%$)}&
\textrm{$\bar{Q}^*(10^{-3})$}&
\textrm{$err$ ($\%$)}\\
\colrule
1/20 & 0.29 & 23.6 & 0.7 & 82.5 \\
1/40 & 0.36 & 5.2  & 3.8 & 5.0  \\
1/60 & 0.38 & $-$  & 4.0 & $-$  \\
\end{tabular}
\end{ruledtabular}
\label{table_grid}
\end{table}

\begin{table}[thbp]
\caption{Tip amplitude ($\Delta x_{\mathrm{tip}}^*$) for different domain height ($H_\mathrm{d}^*$).}
\begin{ruledtabular}
\begin{tabular}{lc}
\textrm{$H_\mathrm{d}^*$}&
\textrm{$\Delta x_{\mathrm{tip}}^*$}\\
\colrule
2 & 1.523  \\
3 & 1.557  \\
4 & 1.560  \\
\end{tabular}
\end{ruledtabular}
\label{table_domainh}
\end{table}

In the present work, uniform grids are used, i.e. $\Delta h = \Delta x =\Delta y = \Delta z$ and $\Delta h = \Delta s$. Several trial calculations are performed to assess the sensitivity of the results to the dimensionless grid size ($\Delta h^* = \Delta h / L$) and the corresponding dimensionless time step ($\Delta t^* = \Delta t / T_b$). The spatial and temporal resolutions, $\Delta h$ ($\Delta x$) and $\Delta t$, are coupled through the constant lattice velocity $c=\Delta x/\Delta t=1$. Therefore, changing the spatial resolution automatically changes the temporal resolution. Figure \ref{fig_gridtest} presents the time histories of the filament tip position in the $x$-direction ($x_{\mathrm{tip}}^*$) and the dimensionless flow rate ($Q^*$) over one beating period, obtained for different $\Delta h^*$ and $\Delta t^*$. The result of $\Delta h^* = 1/40$ agrees well with that of $\Delta h^* = 1/60$. In addition, the tip amplitude ($\Delta x_{\mathrm{tip}}^*$), the time-averaged flow rate ($\bar{Q}^*$), and their corresponding relative errors ($err$) are summarized in Table \ref{table_grid} for different values of $\Delta h^*$. The relative errors decrease significantly when $\Delta h^*$ is reduced from 1/20 to 1/40, and the result for $\Delta h^* = 1/40$ shows good agreement with that for $\Delta h^* = 1/60$. Specifically, the relative errors at $\Delta h^*=1/40$ are only $5.2\%$ for $\Delta x_{\mathrm{tip}}^*$ and $5.0\%$ for $\bar{Q}^*$. Therefore, the grid size of $\Delta h^* = 1/40$ and its corresponding time step of $\Delta t^* = 1.5625 \times 10^{-6}$ are adopted to ensure sufficiently high accuracy while reducing computational cost.

Furthermore, several calculations are conducted to evaluate the effect of the domain height ($H_\mathrm{d}^*$) on filament beating. Table \ref{table_domainh} presents the tip amplitude ($\Delta x_{\mathrm{tip}}^*$) for different $H_\mathrm{d}^*$ ($L_\mathrm{PCL}^*=0.8$, $r_\nu = 10$ and $r_B = 70$). The filament exhibits similar values of $\Delta x_{\mathrm{tip}}^*$ across the tested $H_\mathrm{d}^*$, indicating that the filament beating is insensitive to the domain height. Due to the periodic boundary condition, the present setup effectively represents an infinite array of filaments beating in phase. Although the domain length and width ($L_\mathrm{d}^*=W_\mathrm{d}^*=1$), which determine the filament distribution density, could influence the results through filament coordination, this effect is not explored in the present study, since filament coordination is beyond the current scope.

\section{Results and discussion}
\label{sec:results_and_discussion}
\subsection{Beating pattern of the filament}

\begin{figure}
\centerline{\includegraphics[width=\linewidth]{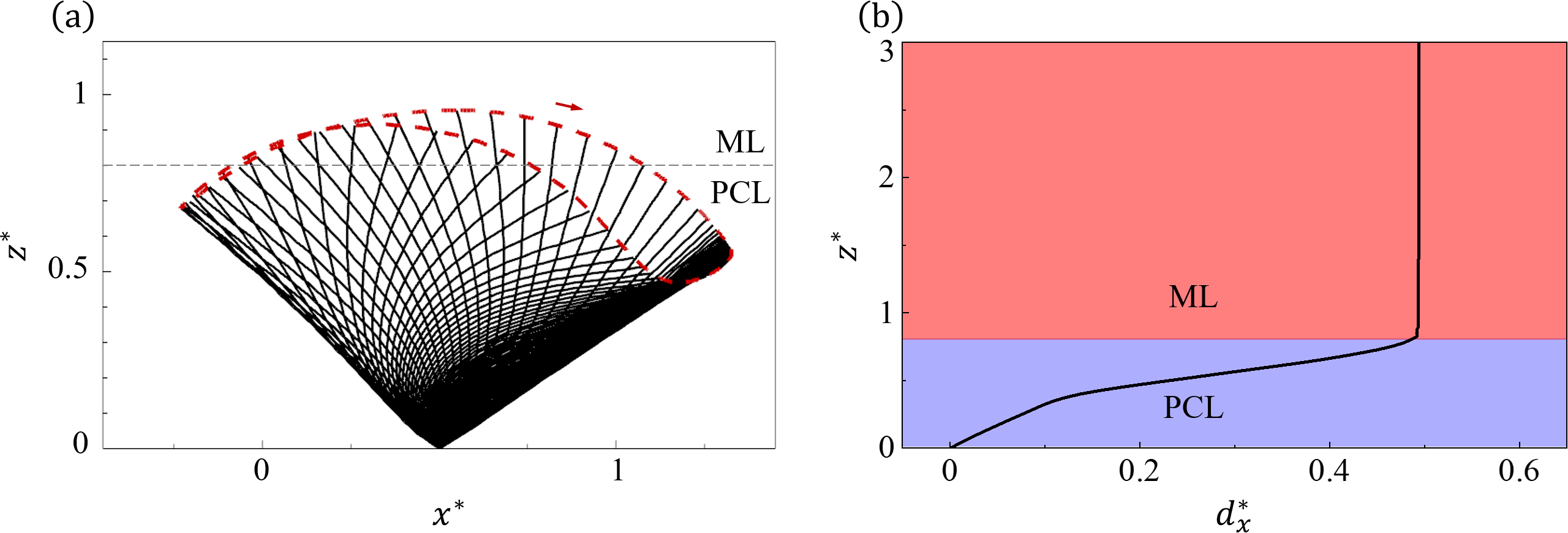}}
\caption{(a) Superimposed instantaneous filament shapes in one beating period. The red dashed line represents the tip trajectory of the filament. The red arrow indicates the direction of tip movement during the power stroke. The grey dashed line represents the initial interface between the PCL and the ML. (b) Fluid displacement in the $x$-direction ($d_x^*$) as a function of $z^*$ in one beating period ($L_\mathrm{PCL}^*=0.8$, $r_\nu = 10$, $r_B = 70$).}
\label{fig_ref_shape}
\end{figure}

We first examine the beating pattern in a representative case ($L_\mathrm{PCL}^* = 0.8$, $r_\nu = 10$, $r_B = 70$), where the flow rate reaches its maximum (Fig. \ref{fig_map}(a)). Figure \ref{fig_ref_shape}(a) presents the superimposed instantaneous filament shapes over one beating period, with a time interval of $6400 \Delta t^*$. The red dashed line indicates the filament tip trajectory, while the grey dashed line marks the initial PCL-ML interface. In the present simulations, the vertical displacement of the PCL-ML interface remains negligible under the action of a single filament. During the power stroke, as the $x$-position of the filament tip increases from its minimum to maximum, the filament exhibits only minor deformation and penetrates the ML for a short duration. In contrast, during the recovery stroke, the filament initially moves slowly beneath the ML with pronounced deformation, followed by a phase of gradual acceleration. The present model captures the essential characteristics of asymmetric ciliary beating, and the resulting beating pattern is generally consistent with experimental observations \citep{gheber1997extraction, chioccioli2019quantitative}. This spatial asymmetry in filament beating results in a net positive flow rate in the $x$-direction. To visualize the fluid transport induced by the filament, the fluid displacement is calculated as $d^*_x (z^*)= \int _0 ^{T_b^*} \bar u^* (z^*,t) dt^*$ on the $x^*=0$ plane, where $z^*$ is the vertical position and $\bar u^*(z^*,t)$ is the spanwise-averaged streamwise velocity. Figure \ref{fig_ref_shape}(b) plots $d^*_x$ as a function of $z^*$, with the PCL region shaded in blue and the ML in red. The PCL acts as a mixing zone characterized by shear flow, whereas the ML serves as a transport zone dominated by nearly uniform flow. These results are consistent with the findings of \citet{ding2014mixing} and \citet{chateau2017transport}.

\begin{figure}
\centerline{\includegraphics[width=\linewidth]{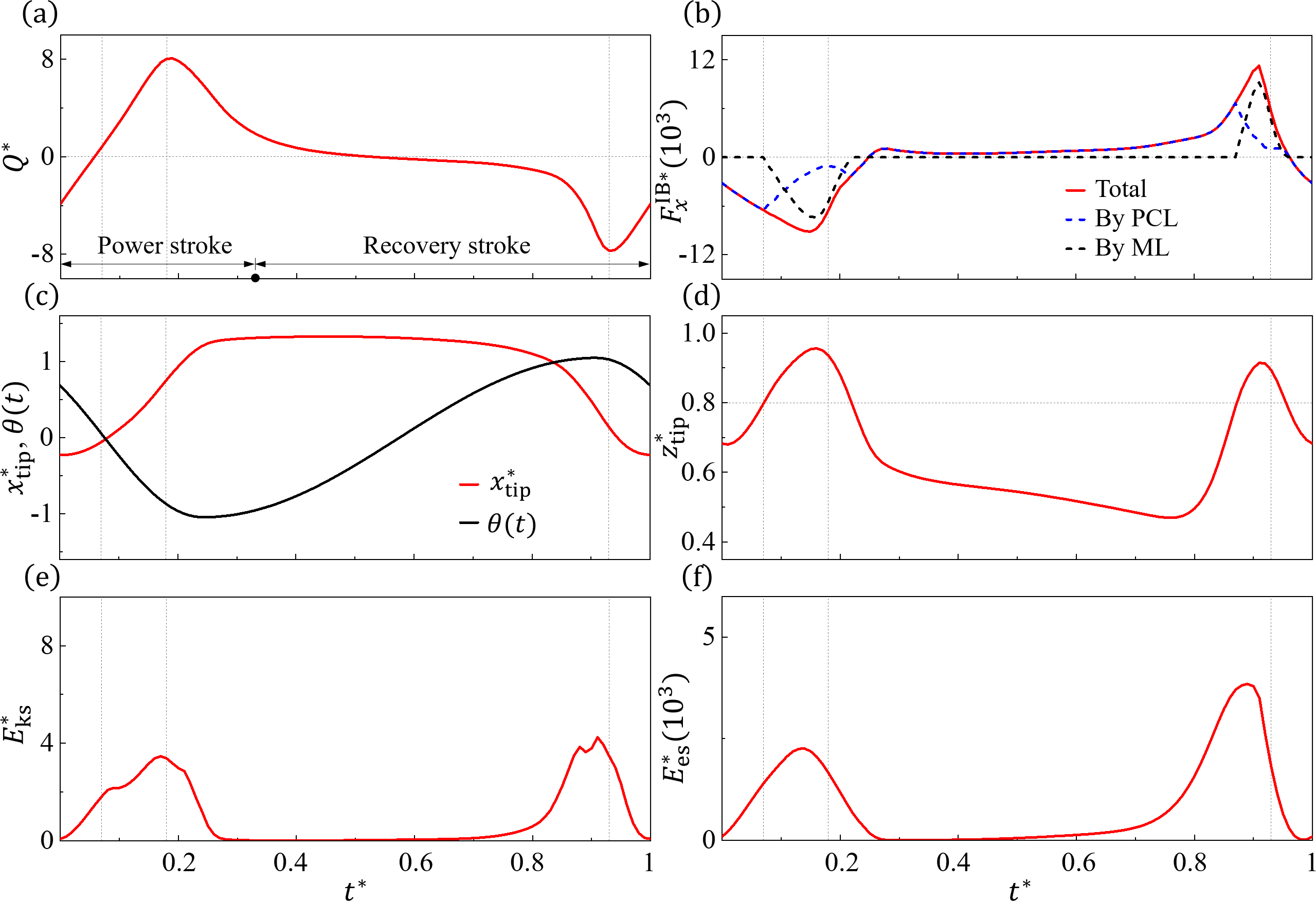}}
\caption{Time histories of (a) the flow rate ($Q^*$), (b) the $x$-component of the fluid force ($F^{\mathrm{IB} *}_x$) and its decomposition into contributions from the PCL and ML, (c) the $x$-position of the filament tip ($x_{\mathrm{tip}}^*$) and basal angle ($\theta (t)$), (d) the $z$-position of the filament tip ($z_{\mathrm{tip}}^*$), (e) the kinetic energy ($E_{\mathrm{ks}}^*$), and (f) the elastic strain energy ($E_{\mathrm{es}}^*$) of the filament ($L_\mathrm{PCL}^*=0.8$, $r_\nu = 10$, $r_B = 70$). The vertical dotted lines correspond to the instants $t^*=$ 0.07, 0.18 and 0.93.}
\label{fig_ref_his}
\vspace{+0.4cm}
\end{figure}

To explore the filament dynamics in detail, the time histories of several quantities in one beating period are shown in Fig. \ref{fig_ref_his}, including the instantaneous flow rate ($Q^*$), the $x$-component of the fluid force ($F^{\mathrm{IB} *}_x$), the $x$-position of the filament tip ($x_{\mathrm{tip}}^*$), the basal angle ($\theta (t)$), the $z$-position of the filament tip ($z_{\mathrm{tip}}^*$), the kinetic energy ($E_{\mathrm{ks}}^*$), and the elastic strain energy ($E_{\mathrm{es}}^*$) of the filament. Here, $t^*=0$ is defined as the instant when $x^*_{\mathrm{tip}}$ reaches its minimum (Fig. \ref{fig_ref_his}(c)), such that the power stroke approximately corresponds to $0 < t^* \leq 1/3$, while the recovery stroke spans $1/3 < t^* \leq 1.0$.

\begin{figure}
\centerline{\includegraphics[width=0.59\linewidth]{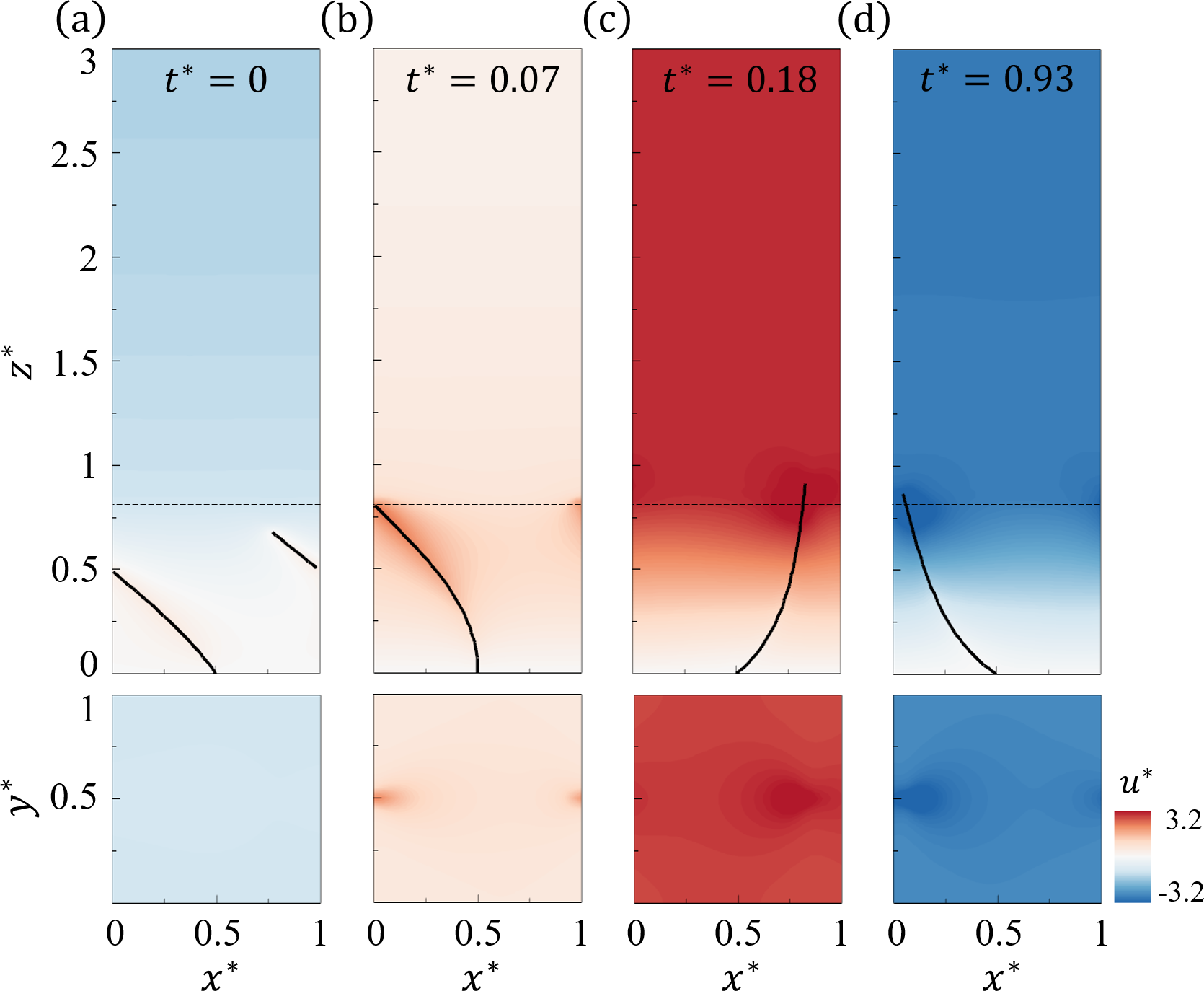}}
\caption{Instantaneous contours of the streamwise flow velocity ($u^*$) in the $y^*=0.5$ and $z^*=0.8$ planes: (a) $t^*=0$, (b) $t^*=0.07$, (c) $t^*=0.18$, and (d) $t^*=0.93$. The grey dashed line represents the initial interface between the ML and the PCL ($L_\mathrm{PCL}^*=0.8$, $r_\nu = 10$, $r_B = 70$).}
\label{fig_ref_velo}
\end{figure}

At $t^*=0$, the flow rate in Fig. \ref{fig_ref_his}(a) is negative because the filament drives the fluid backward during the just-completed recovery stroke. This is visualized by the instantaneous contours of the streamwise velocity ($u^*$) in Fig. \ref{fig_ref_velo}(a). At this instant, the filament tip is located below the ML ($z^*_{\mathrm{tip}} < 0.8$) and reaches its extreme backward position, whereas the rest of the filament has already begun moving forward due to the decrease in the basal angle. Immediately after $t^*=0$, the filament as a whole starts to propel the fluid forward, driven by the continued decrease of the basal angle (Fig. \ref{fig_ref_his}(c)), leading to an increase in both flow rate and filament kinetic energy (Figs. \ref{fig_ref_his}(a) and (e)).

As time progresses to $t^*=0.07$, the filament tip penetrates into the ML (Fig. \ref{fig_ref_velo}(b)), which leads to an increase in the fluid drag contributed by the ML (ML-component of $F^{\mathrm{IB} *}_x$, Fig. \ref{fig_ref_his}(b)). Owing to the resistance of the thick mucus, the growth of the filament kinetic energy stagnates before $t^*=0.11$ (Fig. \ref{fig_ref_his}(e)), even though the basal angle continues to decrease. The ML-component of $\left|F^{\mathrm{IB} *}_x \right|$ rises markedly with increasing penetration depth (Fig. \ref{fig_ref_his}(d)), eventually dominating over the PCL-component. This enhanced drag causes greater filament deformation and a corresponding increase in elastic strain energy (Fig. \ref{fig_ref_his}(f)). Beyond $t^*=0.11$, the filament accelerates noticeably as it has accumulated sufficient elastic strain energy to overcome the drag imposed by the thick mucus. The results indicate that the ML-component of $F^{\mathrm{IB} *}_x$ significantly influences filament dynamics, even though the filament only slightly penetrates the ML for a short duration. The time-averaged total fluid drag and its ML-component are $\bar{F}^{\mathrm{IB} *}_{x}=2.87\times10^3$ and $\bar{F}^{\mathrm{IB} *}_{x\mathrm{ML}}=1.04\times10^3$, respectively. The corresponding ratio ($r_F=\bar{F}^{\mathrm{IB} *}_{x\mathrm{ML}}/\bar{F}^{\mathrm{IB} *}_{x}$) is approximately 0.36, and can increase up to 0.7 as $L_\mathrm{PCL}^*$ and $r_\nu$ are varied (Fig. \ref{fig_map}(d)).

At $t^*=0.18$, both the flow rate and the filament kinetic energy reach their local maxima. Owing to the low inertial effect, the mucus flow is effectively guided by the filament, as shown in Fig. \ref{fig_ref_velo}(c), where the positive flow occupies nearly the entire domain. At $t^*=0.24$, the basal angle reaches its minimum, corresponding to a decrease of the bending stiffness to $B_\mathrm{min}^*$. Under these conditions, the free portion of the softened filament is unable to respond in time to the imposed base rotation due to fluid drag, leading to slow filament motion and a pronounced phase shift between $\theta(t)$ and $x_{\mathrm{tip}}^*$ (Fig. \ref{fig_ref_his}(c)). As a result, the fluid drag, filament kinetic energy, and elastic strain energy decrease substantially, approaching zero (Figs. \ref{fig_ref_his}(b), (e) and (f). Beyond $t^*=0.24$, the filament gradually stiffens. During $t^*=0.24-0.7$, the bending stiffness increases slowly due to the power index $n$, leading to gradual rises in the filament's elastic strain and kinetic energies. The flow rate also decreases slowly owing to the curved filament shape. In the subsequent phase ($0.7 \leq t^* \leq 0.91$), the bending stiffness recovers more rapidly, resulting in a pronounced increase in elastic strain energy. During this interval, the filament's elastic force overcomes fluid drag, causing a rapid rise in kinetic energy and a sharp drop in the flow rate.

As time progresses to $t^*=0.93$, the flow rate reaches its minimum, with negative flow occupying nearly the entire area (Fig. \ref{fig_ref_velo}(d)). At this instant, the basal angle attains its maximum, while the curved filament still retains relatively high elastic strain energy. Subsequently, the filament tip and base move in opposite directions, leading to a decrease in kinetic energy and an increase in flow rate. In summary, the filament drives fluid forward during the power stroke and backward during the recovery stroke, with the time-averaged flow rate remaining positive due to the spatial asymmetry of its beating.

\subsection{Effects of PCL thickness and viscosity ratio}
\subsubsection{Overview of results}

\begin{figure}
\centerline{\includegraphics[width=0.98\linewidth]{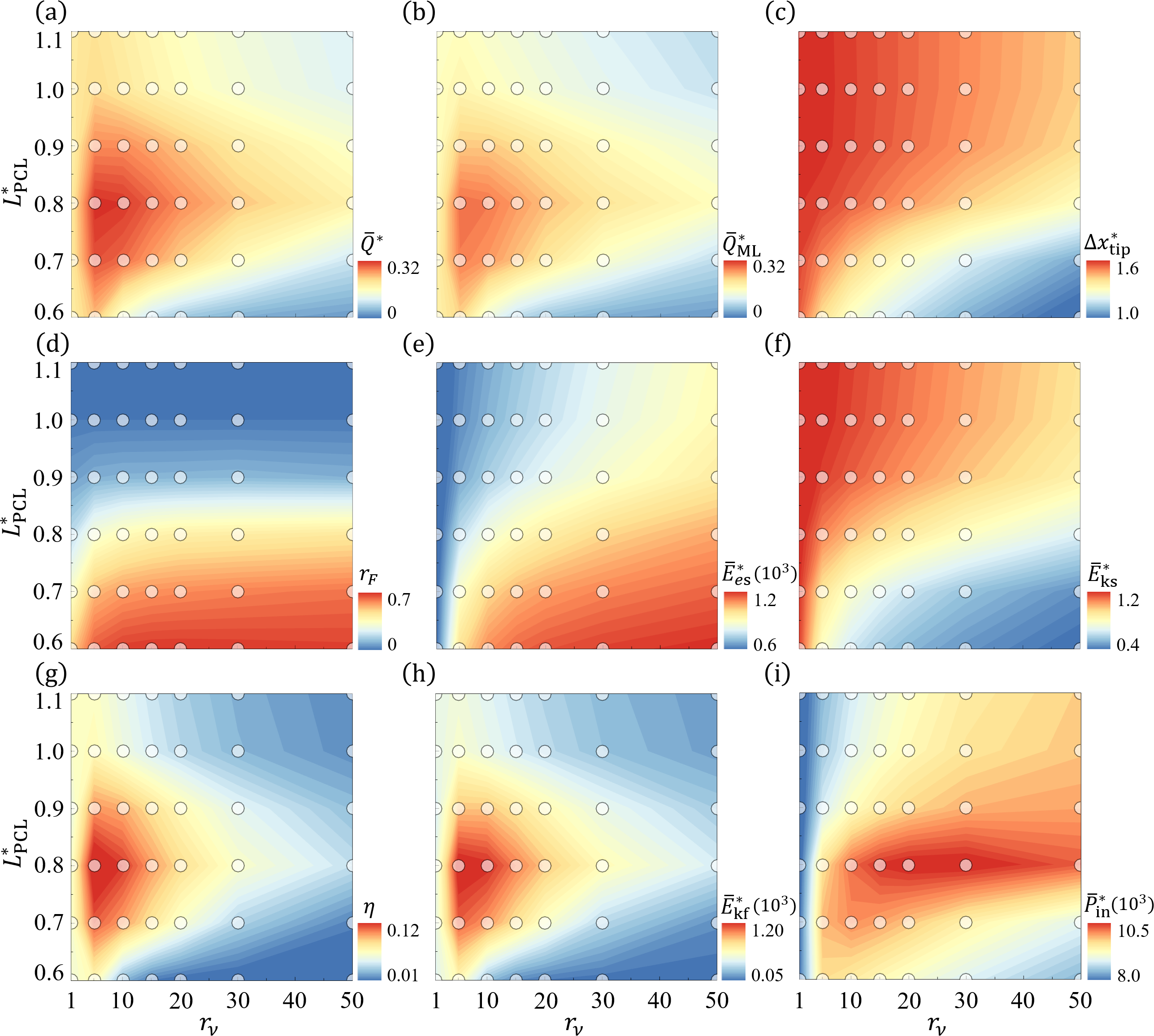}}
\caption{Interpolated contours depending on $L_\mathrm{PCL}^*$ and $r_\nu$: (a) the time-averaged flow rate ($\bar{Q}^*$), (b) the time-averaged flow rate of the ML ($\bar{Q}^*_{\mathrm{ML}}$), (c) the beating amplitude of the filament tip ($\Delta x_{\mathrm{tip}}^*$) in the $x$-direction, (d) the ratio ($r_F$) between the drag caused by the ML and the total drag, (e) the time-averaged elastic strain energy ($\bar{E}_{\mathrm{es}}^*$) of the filament, (f) the time-averaged kinetic energy ($\bar{E}_{\mathrm{ks}}^*$) of the filament, (g) the transport efficiency ($\eta$), (h) the time-averaged effective kinetic energy ($\bar{E}_{\mathrm{kf}}^*$) of the fluid, and (i) the time-averaged input power ($\bar{P}_{\mathrm{in}}^*$) of the filament ($r_B=70$). The circular symbols are data points.}
\label{fig_map}
\end{figure}

As discussed above, the filament motion is affected when it penetrates into the ML. Both the PCL thickness ($L_\mathrm{PCL}^*$) and the viscosity ratio ($r_\nu$) directly influence the filament dynamics. Figure \ref{fig_map} provides an overview of the combined effects of $L_\mathrm{PCL}^*$ and $r_\nu$, with several interpolated contours shown in the ($L_\mathrm{PCL}^*$, $r_\nu$) parameter space ($r_B=70$).

On the whole, the predicted time-averaged flow rate ($\bar Q^*$) across the plane $x^*=0$ ranges from 0.024 to 0.32, corresponding to dimensional flow rates of $2.74 \times 10^{-7}$ to $3.65 \times 10^{-6}$ $\mathrm{\mu L/s}$. These values represent the transport generated by a single filament only. The dimensional flow velocity obtained in the present study varies between 1.427 and 19.01 $\mathrm{\mu m/s}$, which is close to the experimental measurements of \citet{ueno2012mouse} (approximately 10-20 $\mathrm{\mu m/s}$), but considerably lower than the value reported by \citet{morgan2004scintigraphic} (approximately 178.3 $\mathrm{\mu m/s}$). This discrepancy likely arises from the simplified setup adopted here, in which only a single filament is modeled. Owing to the periodic boundary condition, this corresponds to an infinite array of filaments beating in phase. In contrast, in biological systems, phase differences between neighboring cilia generate metachronal waves that markedly enhance flow velocity. Therefore, the flow velocity values obtained in the present single-filament model should be regarded as a lower bound compared to those achievable in physiological cilia arrays. The streamwise tip amplitude of the filament ($\Delta x^*_{\mathrm{tip}}$) predicted by the present model ranges from 0.953 to 1.614, corresponding to dimensional values of 7.624 to 12.912 $\mathrm{\mu m}$. These are close to the experimental estimate of approximately 9.4 $\mathrm{\mu m}$ \citep{gheber1997extraction,chioccioli2019quantitative}, further confirming the ability of the present model to reproduce the essential beating pattern of the cilium.

The contours of the time-averaged total flow rate ($\bar Q^*$) and the time-averaged flow rate within the ML ($\bar Q^*_{\mathrm{ML}}$) exhibit similar trends, as shown in Figs. \ref{fig_map}(a) and (b), indicating that the total flow rate is predominantly determined by the ML. For instance, the ratio $\bar Q^*_{\mathrm{ML}} / \bar Q^*$ is approximately 0.88 when $L_\mathrm{PCL}^* = 0.8$ and $r_{\nu} = 10$. Both $\bar Q^*$ and $\bar Q^*_{\mathrm{ML}}$ attain their maximum values at $L_\mathrm{PCL}^* = 0.8$. This result is consistent with the dimensionless PCL thickness (0.75$-$0.875) estimated from experimental data, as discussed in Section \ref{sec:computational_model}. It also agrees with the observation of \citet{jayathilake2015numerical} that the filament slightly penetrates into the ML to maximize fluid transport. Furthermore, the viscosity ratio that maximizes fluid transport is found to lie between 5 and 10 for different $L_\mathrm{PCL}^*$ values. This range is consistent with the physiological viscosity ratio of healthy ML and PCL (5$-$50), as discussed in Section \ref{sec:computational_model}, though it is slightly lower than the optimal $r_{\nu}$ of 10$-$20 reported in the numerical study of \citet{chatelin2016parametric}. This difference can be attributed to the two-way coupling system adopted here, which incorporates fluid feedback on the filament. As $r_{\nu}$ increases, the reduction in filament tip amplitude diminishes propulsion efficiency, thereby shifting the optimal $r_{\nu}$ toward lower values. We will discuss the details later.

Figure \ref{fig_map}(c) shows the contour of the tip amplitude ($\Delta x^*_{\mathrm{tip}}$). With increasing $r_{\nu}$ and decreasing $L_\mathrm{PCL}^*$, $\Delta x^*_{\mathrm{tip}}$ decreases. The maximum flow rate is achieved at an intermediate tip amplitude. Notably, the largest tip amplitude occurs when the filament beats entirely within the PCL, where it exerts minimal force on the mucus layer. Figure \ref{fig_map}(d) illustrates the contour of the drag ratio ($r_F$), defined as the drag exerted by the ML relative to the total drag, which exhibits an opposite trend to $\Delta x^*_{\mathrm{tip}}$. As $L_\mathrm{PCL}^*$ decreases and $r_{\nu}$ increases, the ML contribution to the total drag rises, leading to reduced tip amplitudes and greater filament deformation under the influence of the thicker mucus layer. This explains the variations in $\Delta x^*_{\mathrm{tip}}$ and $\bar E^*_{\mathrm{es}}$ (time-averaged elastic strain energy) observed in Figs. \ref{fig_map}(c) and (e). The effect of $L_\mathrm{PCL}^*$ is particularly significant: the contour of $r_F$ can be broadly divided into a high-$r_F$ region ($L_\mathrm{PCL}^* < 0.8$) and a low-$r_F$ region ($L_\mathrm{PCL}^* > 0.8$). In addition, the contours of $\Delta x^*_{\mathrm{tip}}$ and $\bar E^*_{\mathrm{ks}}$ (time-averaged kinetic energy of the filament) exhibit similar patterns (Figs. \ref{fig_map}(c) and (f)), as the tip velocity increases with tip amplitude at a fixed frequency, resulting in larger $\bar E^*_{\mathrm{ks}}$.

The contour of the transport efficiency ($\eta$, Fig. \ref{fig_map}(g)) resembles that of the flow rate ($\bar Q^*$, Fig. \ref{fig_map}(a)), indicating that high efficiency and high flow rate occur simultaneously. This behavior is primarily governed by the time-averaged effective kinetic energy of the fluid ($\bar{E}_{\mathrm{kf}}^*$, Fig. \ref{fig_map}(h)), which is comparable in magnitude to the filament's elastic strain energy ($\bar{E}_{\mathrm{es}}^*$, Fig. \ref{fig_map}(e)) and substantially larger than its kinetic energy ($\bar{E}_{\mathrm{ks}}^*$, Fig. \ref{fig_map}(f)). Since both $\bar Q^*$ and $\bar{E}_{\mathrm{kf}}^*$ quantify the forward propulsion of the fluid, their contours exhibit similar patterns, with $\bar{E}_{\mathrm{kf}}^*$ being proportional to $\bar Q^*$. Note that the input power in the present system is not prescribed and cannot be externally controlled; rather, it emerges as the outcome of the dynamic balance between filament motion and fluid motion. In Fig. \ref{fig_map}(i), elevated values of the time-averaged input power ($\bar{P}_{\mathrm{in}}^*$) are observed for $L_\mathrm{PCL}^*=0.8$ and $r_{\nu}>10$; however, most of the additional input power is converted into elastic strain energy and spent offsetting the negative flow produced during the recovery stroke, rather than enhancing net propulsion.

The above results underscore the strong influence of $L_\mathrm{PCL}^*$ and $r_\nu$ on fluid transport and filament beating dynamics. A key question naturally arises: although a larger tip amplitude is generally expected to enhance fluid propulsion, why do the maximum flow rate and transport efficiency not occur at the largest amplitudes? This issue will be examined in detail below.

\subsubsection{Effects of PCL thickness}

\begin{figure}
\centerline{\includegraphics[width=\linewidth]{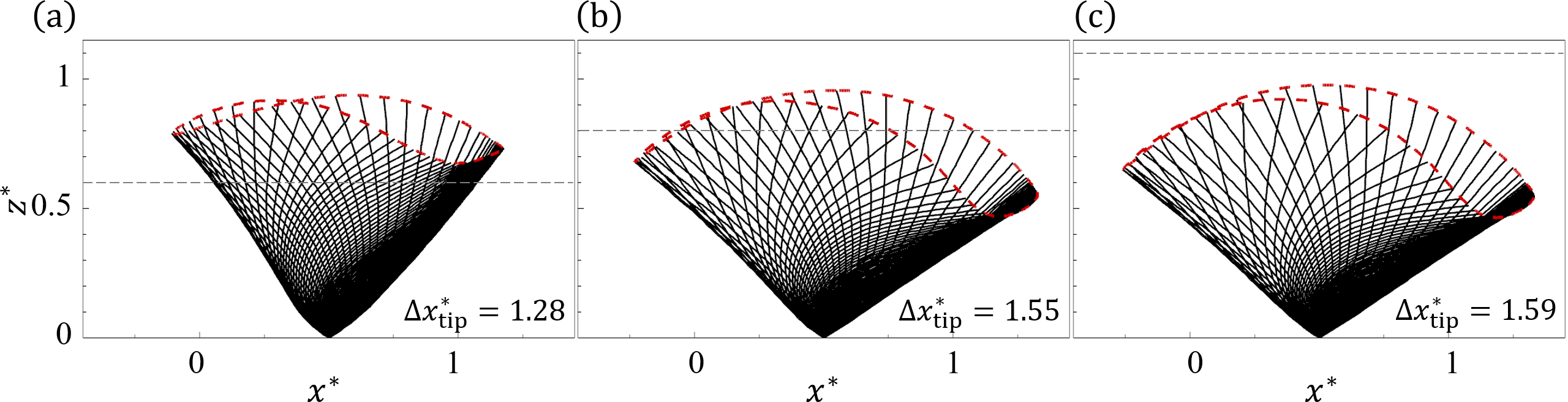}}
\caption{Superimposed instantaneous filament shapes in one beating period for (a) $L_\mathrm{PCL}^*=0.6$, (b) $L_\mathrm{PCL}^*=0.8$ and (c) $L_\mathrm{PCL}^*=1.1$ ($r_\nu=10$, $r_B=70$).}
\label{fig_lpcl_shape}
\end{figure}

We first investigate the effect of PCL thickness, which strongly influences both fluid transport and the filament's beating pattern. Figure \ref{fig_lpcl_shape} presents the superimposed instantaneous filament shapes for different $L_\mathrm{PCL}^*$ ($r_\nu=10$, $r_B=70$). The beating amplitude increases noticeably with increasing $L_\mathrm{PCL}^*$. For $L_\mathrm{PCL}^* = 0.6$ (Fig. \ref{fig_lpcl_shape}(a)), the filament tip remains entirely within the ML throughout the cycle. At $L_\mathrm{PCL}^* = 0.8$ (Fig. \ref{fig_lpcl_shape}(b)), the filament penetrates the ML briefly, and the larger spacing between adjacent positions indicates an increase in kinetic energy. For $L_\mathrm{PCL}^* = 1.1$ (Fig. \ref{fig_lpcl_shape}(c)), the filament remains fully below the ML, and the beating amplitude reaches its maximum owing to the absence of ML-induced resistance.

\begin{table}[b]
\caption{Time-averaged flow rate ($\bar{Q}^*$), input power ($\bar{P}_{\mathrm{in}}^*$), effective kinetic energy of the fluid ($\bar{E}_{\mathrm{kf}}^*$), elastic strain energy of the filament ($\bar{E}_{\mathrm{es}}^*$), total fluid drag ($\bar{F}^{\mathrm{IB} *}_x$), ML-component of the fluid drag ($\bar{F}^{\mathrm{IB} *}_{x\mathrm{ML}}$) for different $L_\mathrm{PCL}^*$ ($r_\nu = 10$, $r_B = 70$).}
\begin{ruledtabular}
\begin{tabular}{ccccccc}
\textrm{$L_\mathrm{PCL}^*$}&
\textrm{$\bar{Q}^*$}&
\textrm{$\bar{P}_{\mathrm{in}}^*$} ($10^3$)&
\textrm{$\bar{E}_{\mathrm{kf}}^*$} ($10^3$)&
\textrm{$\bar{E}_{\mathrm{es}}^*$} ($10^3$)&
\textrm{$\bar{F}^{\mathrm{IB} *}_x$} ($10^3$)&
\textrm{$\bar{F}^{\mathrm{IB} *}_{x\mathrm{ML}}$} ($10^3$)\\
\colrule
0.6 & 0.141 & 9.52 & 0.23 & 1.10 & 3.49 & 2.42 \\
0.8 & 0.320 & 10.27 & 1.23 & 0.85 & 2.87 & 1.04 \\
1.1 & 0.191 & 8.75 & 0.44 & 0.66 & 2.52 & 0 \\
\end{tabular}
\end{ruledtabular}
\label{table_avevalue_lpcl}
\end{table}

\begin{figure}
\centerline{\includegraphics[width=\linewidth]{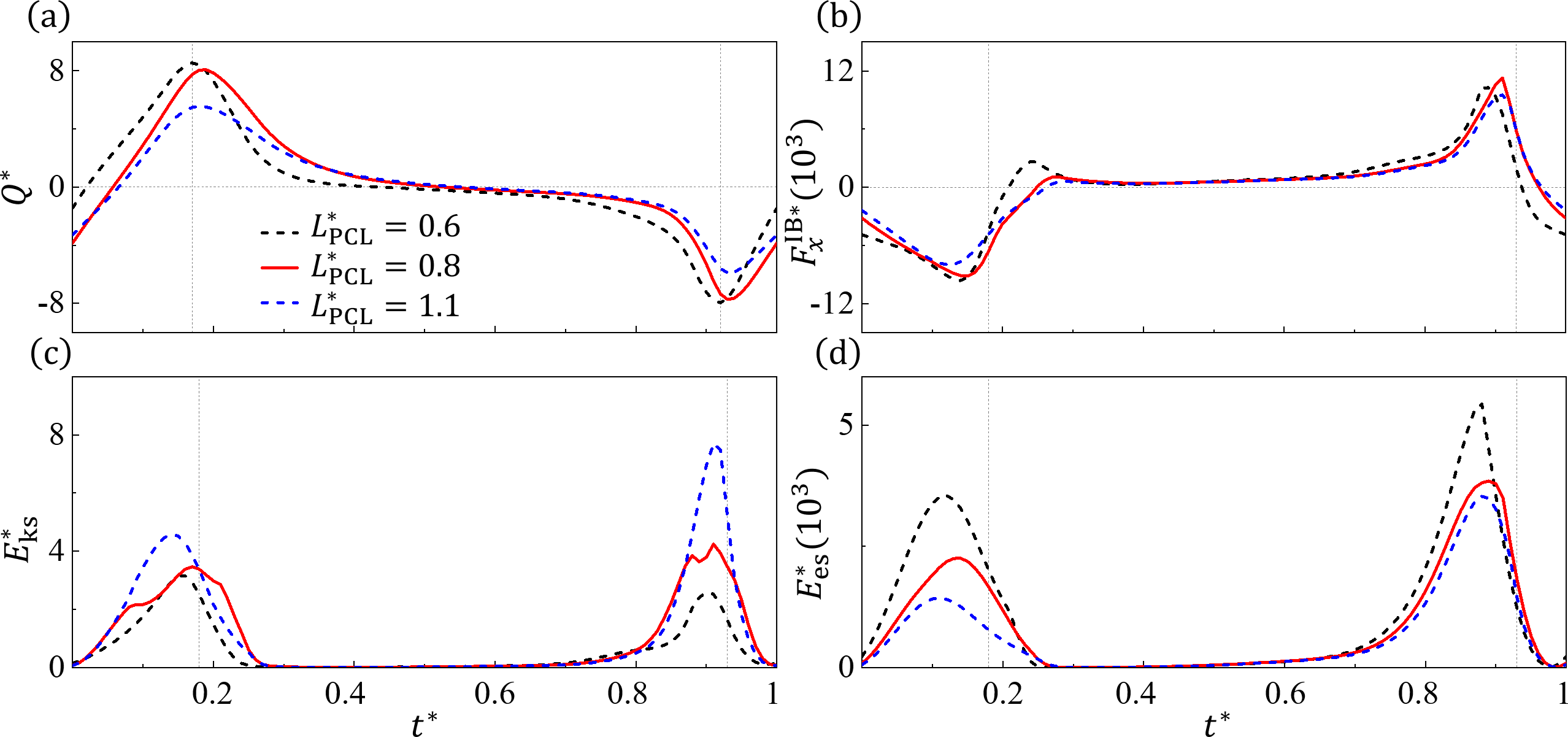}}
\caption{Time histories of (a) $Q^*$, (b) $F^{\mathrm{IB} *}_x$, (c) $E_{\mathrm{ks}}^*$ and (d) $E_{\mathrm{es}}^*$ for different $L_\mathrm{PCL}^*$ ($r_\nu=10$, $r_B=70$). The vertical dotted lines correspond to the instants $t^*=$ 0.17 and 0.92.}
\label{fig_lpcl_his}
\end{figure}

\begin{figure}
\centerline{\includegraphics[width=\linewidth]{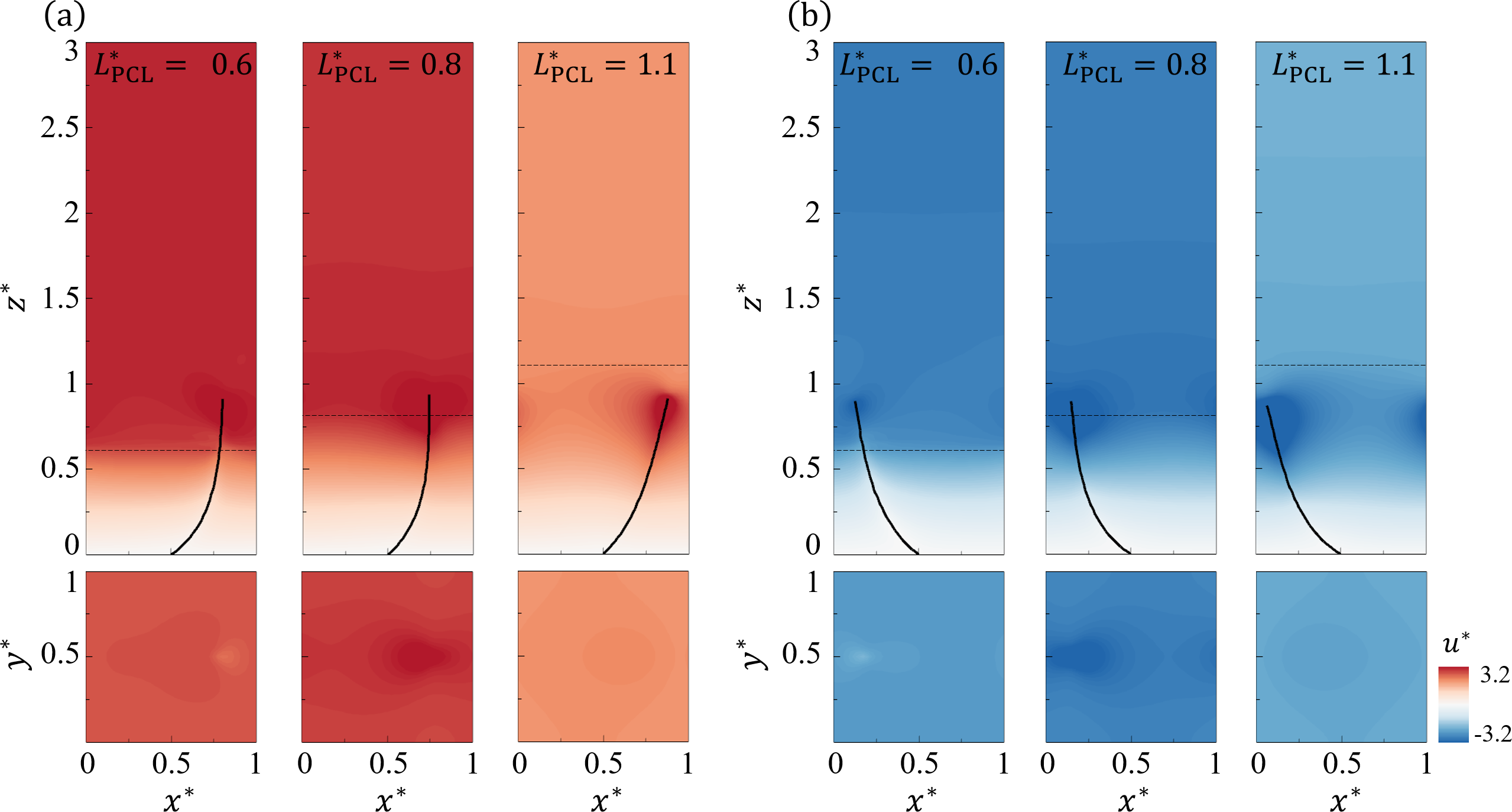}}
\caption{Instantaneous contours of $u^*$ in the $y^*=0.5$ and $z^*=L_\mathrm{PCL}^*$ planes for different $L_\mathrm{PCL}^*$: (a) $t^*=0.17$ and (b) $t^*=0.92$ ($r_\nu=10$, $r_B=70$).}
\label{fig_lpcl_velo}
\end{figure}

To quantify the resistance imposed by the ML, Table \ref{table_avevalue_lpcl} lists the time-averaged total fluid drag ($\bar{F}^{\mathrm{IB} *}_x$) and its ML-component ($\bar{F}^{\mathrm{IB} *}_{x\mathrm{ML}}$). Both $\bar{F}^{\mathrm{IB} }_x$ and $\bar{F}^{\mathrm{IB} *}_{x\mathrm{ML}}$ decrease markedly as $L_\mathrm{PCL}^*$ increases, leading to a reduction in the drag ratio ($r_F$) from 0.69 to 0 (Fig. \ref{fig_map}(d)). The ML-component drag acts primarily near the filament tip. With bending stiffness ratio fixed, a larger $\bar{F}^{\mathrm{IB} *}_{x\mathrm{ML}}$ produces greater filament deformation, reflected in higher elastic strain energy (Table \ref{table_avevalue_lpcl} and Fig. \ref{fig_lpcl_his}(d)). This deformation allows the filament to generate stronger elastic forces that counterbalance the fluid drag. Consequently, the tip amplitude is reduced, which in turn lowers the filament's kinetic energy (Fig. \ref{fig_lpcl_his}(c)).

A larger tip amplitude (and thus higher filament's kinetic energy) does not always enhance flow propulsion in the present two-phase system. As shown in Figs. \ref{fig_lpcl_his}(a) and \ref{fig_lpcl_velo}, a strong positive flow rate is produced at $t^* \approx 0.17$, followed by a large-magnitude negative flow rate at $t^* \approx 0.92$. These instantaneous peaks are attenuated as $L_\mathrm{PCL}^*$ increases (tip amplitude increases), underscoring the critical role of filament penetration into the ML. For $L_\mathrm{PCL}^*=0.6$, the penetration depth is largest, which promotes mucus transport. Owing to the higher viscosity of the ML compared to the PCL, the enhanced viscous diffusion of momentum allows the mucus flow to follow the filament motion effectively, as evident in Fig. \ref{fig_lpcl_velo}, where positive/negative flow occupies nearly the entire domain. However, this also causes the flow rate to decay more rapidly after reaching its positive peak and leads to a larger-magnitude negative flow rate than in the other cases. For $L_\mathrm{PCL}^*=0.8$, the filament penetrates the ML only briefly, and both the positive and negative peak flow rates are only slightly reduced due to the increases in the tip amplitude and the filament's kinetic energy. For $L_\mathrm{PCL}^*=1.1$, the filament remains entirely in the PCL, where the weaker viscous diffusion of momentum substantially attenuates the peak flow rates, due to the fact that the tip amplitude only slightly increases compared to that for $L_\mathrm{PCL}^*=0.8$. The filament primarily influences the fluid motion in its immediate vicinity, as shown in Fig. \ref{fig_lpcl_velo}. Since the net flow depends on both the positive contribution from the power stroke and the negative contribution from the recovery stroke, these were quantified by integrating the positive and negative flow rates over one cycle: $Q^{*}_{+}=\int Q^* dt^*$ (for $Q^* > 0$) and $Q^{*}_{-}=\int Q^* dt^*$ (for $Q^* < 0$). The results show $Q^{*}_{+}=1.318, 1.413$, and $1.076$ and $Q^{*}_{-}=$ -1.177, -1.093, and -0.885 for $L_\mathrm{PCL}^*=0.6, 0.8$, and $1.1$, respectively. Thus, at $L_\mathrm{PCL}^*=0.8$, the filament produces the largest $Q^{*}_{+}$ together with a moderate $Q^{*}_{-}$, yielding the highest net flow rate.

Compared with the cases of $L_\mathrm{PCL}^* = 0.6$ and $1.1$, the filament at $L_\mathrm{PCL}^* = 0.8$ consumes the largest input power, a greater fraction of which is converted into the effective kinetic energy of the fluid (Table \ref{table_avevalue_lpcl}). This leads to the highest transport efficiency. In contrast, for $L_\mathrm{PCL}^* = 0.6$ and $1.1$, most of the input power is dissipated into elastic strain energy or expended in counteracting the negative flow generated during the recovery stroke.

In summary, the net flow rate results from the combined effects of two mechanisms: the force balance between fluid drag and elastic force (drag-elastic force balance), and the viscous diffusion of momentum. The former governs the tip amplitude and filament deformation, thereby reflecting the capacity to propel fluid, while the latter dictates the extent to which the flow responds to filament motion. These two mechanisms act in competition as $L_\mathrm{PCL}^*$ varies. Optimal transport is achieved when the filament slightly penetrates the ML, allowing for relatively low fluid drag while maintaining effective interaction between the filament and the ML.

\subsubsection{Effects of viscosity ratio}

\begin{figure}
\centerline{\includegraphics[width=\linewidth]{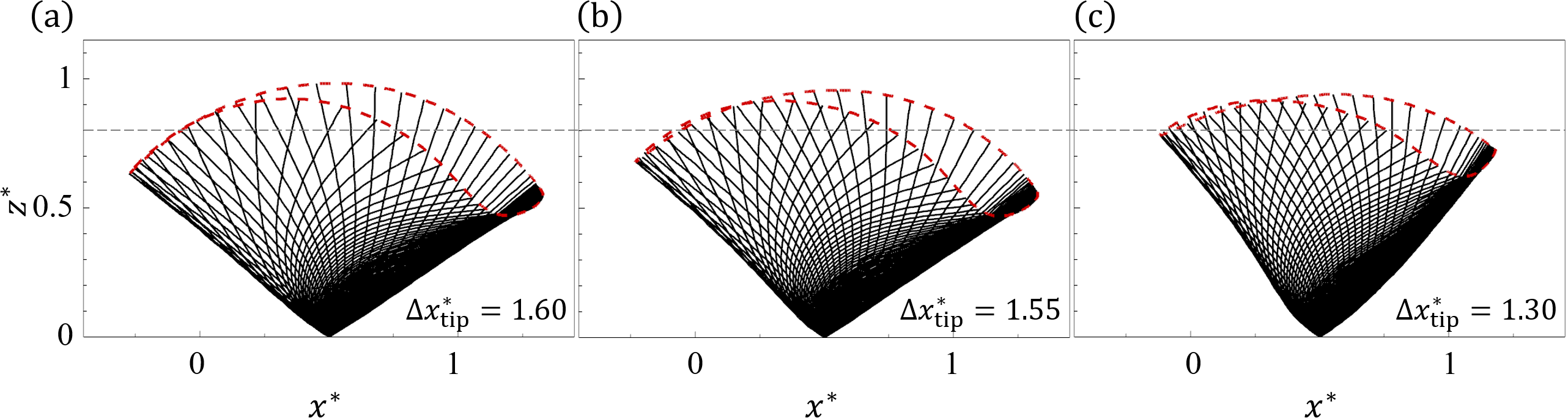}}
\caption{Superimposed instantaneous filament shapes in one beating period for (a) $r_\nu=1$, (b) $r_\nu=10$, and (c) $r_\nu=50$ ($L_\mathrm{PCL}^*=0.8$, $r_B=70$).}
\label{fig_rnu1_shape}
\end{figure}

\begin{figure}
\centerline{\includegraphics[width=\linewidth]{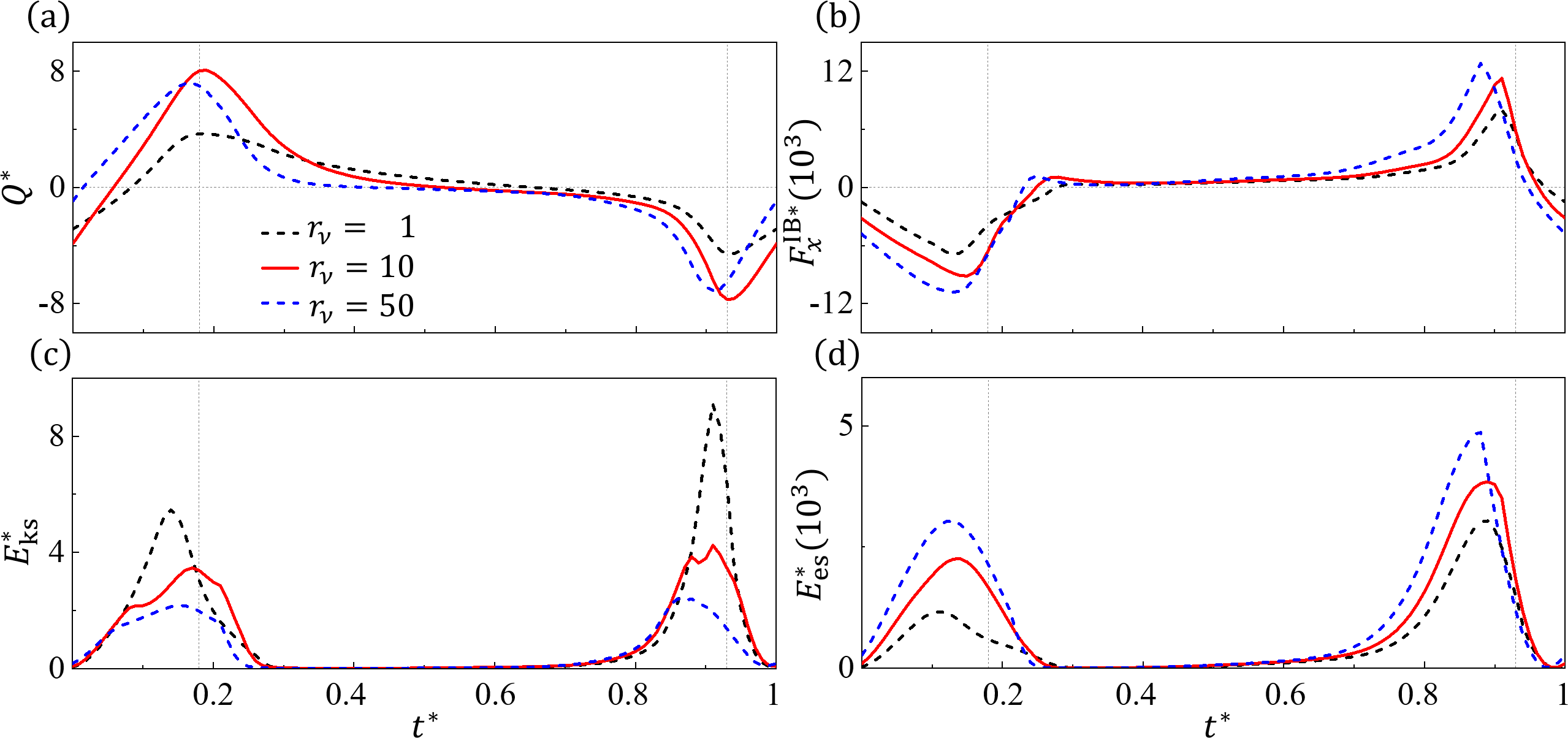}}
\caption{Time histories of (a) $Q^*$, (b) $F^{\mathrm{IB} *}_x$, (c) $E_{\mathrm{ks}}^*$, and (d) $E_{\mathrm{es}}^*$ for different $r_\nu$ ($L_\mathrm{PCL}^*=0.8$, $r_B=70$). The vertical dotted lines correspond to the instants $t^*=$ 0.18 and 0.93.}
\label{fig_rnu1_his}
\end{figure}

Here we further examine the effect of $r_{\nu}$. When $r_{\nu}=1$, the two-phase flow reduces to a single-phase flow. The superimposed instantaneous filament shapes for different $r_{\nu}$ are shown in Fig. \ref{fig_rnu1_shape} ($L_\mathrm{PCL}^*=0.8$, $r_B=70$). As $r_{\nu}$ increases from 1 to 50, a clear decrease in the tip amplitude is observed, which corresponds to increases in both the total fluid drag and the ML-component fluid drag (Table \ref{table_avevalue_rnu} and Fig. \ref{fig_rnu1_his}). Consequently, the contribution of the ML to the overall fluid drag ($r_F$) rises from 0.17 to 0.44 (Fig. \ref{fig_map}(d)). This trend is consistent with that observed for decreasing $L_\mathrm{PCL}^*$ (Table \ref{table_avevalue_lpcl} and Fig. \ref{fig_lpcl_shape}). The reduced tip amplitude indicates stronger filament deformation, accompanied by higher elastic strain energy and lower kinetic energy of the filament (Table \ref{table_avevalue_rnu}; Figs. \ref{fig_rnu1_his}(c) and (d)).

\begin{table}[b]
\caption{Time-averaged flow rate ($\bar{Q}^*$), input power ($\bar{P}_{\mathrm{in}}^*$), effective kinetic energy of the fluid ($\bar{E}_{\mathrm{kf}}^*$), elastic strain energy of the filament ($\bar{E}_{\mathrm{es}}^*$), total fluid drag ($\bar{F}^{\mathrm{IB} *}_x$), ML-component of the fluid drag ($\bar{F}^{\mathrm{IB} *}_{x\mathrm{ML}}$) for different $r_\nu$ ($L_\mathrm{PCL}^*=0.8$, $r_B = 70$).}
\begin{ruledtabular}
\begin{tabular}{ccccccc}
\textrm{$r_\nu$}&
\textrm{$\bar{Q}^*$}&
\textrm{$\bar{P}_{\mathrm{in}}^*$} ($10^3$)&
\textrm{$\bar{E}_{\mathrm{kf}}^*$} ($10^3$)&
\textrm{$\bar{E}_{\mathrm{es}}^*$} ($10^3$)&
\textrm{$\bar{F}^{\mathrm{IB} *}_x$} ($10^3$)&
\textrm{$\bar{F}^{\mathrm{IB} *}_{x\mathrm{ML}}$} ($10^3$)\\
\colrule
1 & 0.196 & 7.80 & 0.47 & 0.55 & 2.07 & 0.36 \\
10 & 0.320 & 10.27 & 1.23 & 0.85 & 2.87 & 1.04 \\
50 & 0.183 & 10.34 & 0.40 & 1.07 & 3.80 & 1.66 \\
\end{tabular}
\end{ruledtabular}
\label{table_avevalue_rnu}
\end{table}

\begin{figure}
\centerline{\includegraphics[width=\linewidth]{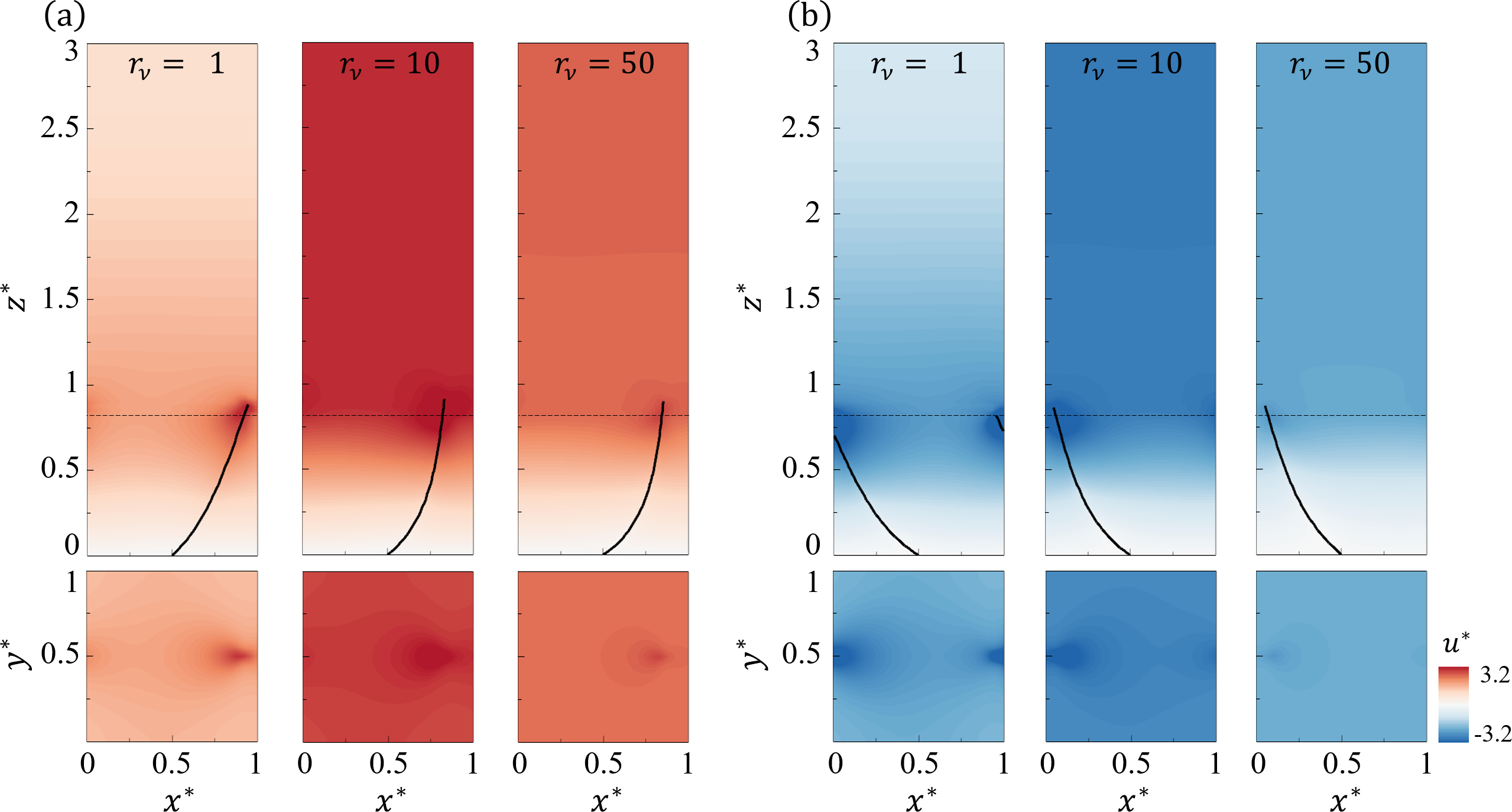}}
\caption{Instantaneous contours of $u^*$ in the $y^*=0.5$ and $z^*=0.8$ planes for different $r_\nu$: (a) $t^*=0.18$ and (b) $t^*=0.93$ ($L_\mathrm{PCL}^*=0.8$, $r_B=70$).}
\label{fig_rnu1_velo}
\end{figure}

As $r_\nu$ varies, the net flow rate continues to be governed by the competition between the drag-elastic force balance and the viscous diffusion of momentum. For $r_\nu=1$, the ML viscosity is low, resulting in weak viscous diffusion. Consequently, the filament influences the fluid motion only in its immediate vicinity (Fig. \ref{fig_rnu1_velo}), and the positive/negative flow rates during the power/recovery strokes are attenuated (Fig. \ref{fig_rnu1_his}(a)), despite the high tip amplitude and kinetic energy of the filament. For $r_\nu=10$, the enhanced viscous diffusion, combined with the slightly reduced tip amplitude, produces a higher instantaneous flow rate. When $r_\nu$ is further increased to 50, the strong fluid drag substantially reduces the tip amplitude (i.e. weakens the propulsive capacity of the filament), thereby lowering the instantaneous flow rate. Quantitatively, the filament at $r_\nu=10$ generates a much higher positive flow ($Q^{*}_{+}=1.413$) than at $r_\nu=1$ (0.897) and $r_\nu=50$ (1.162), while its negative flow ($Q^{*}_{-}=$ -1.093) is slightly lower than that at $r_\nu=1$ (-0.701) and $r_\nu=50$ (-0.978). Thus, the increase in $Q^{*}_{+}$ outweighs the decrease in $Q^{*}_{-}$, yielding a higher net flow rate.

The input power increases by 31.6$\%$ as $r_\nu$ rises from 1 to 10, and this additional input is efficiently converted into fluid kinetic energy (increases by 161.7$\%$, Table \ref{table_avevalue_rnu}), thereby enhancing transport efficiency. However, when $r_\nu$ further increases to 50, the input power remains nearly unchanged while less energy is transferred into fluid kinetic energy, leading to a marked decline in transport efficiency. Under this condition, a larger fraction of energy is dissipated as elastic strain energy and expended in counteracting the negative flow rate.

For the beating patterns shown in Figs. \ref{fig_lpcl_shape}(a) and \ref{fig_rnu1_shape}(c), the asymmetric nature of the beat is preserved even though the tip amplitude is small. In the extreme case with $L_\mathrm{PCL}^*=0.6$ and $r_{\nu}=50$, however, the beating asymmetry is almost lost. Such a beating pattern with low tip amplitude is inefficient for fluid transport, and the time-averaged flow rate approaches zero. The present model therefore predicts one possible scenario for diseased ASL, such as in cystic fibrosis and chronic obstructive pulmonary disease, where an excessively thick ML envelops the cilia and restricts the tip motion.

\subsection{Effects of bending stiffness ratio}

\begin{figure}
\centerline{\includegraphics[width=\linewidth]{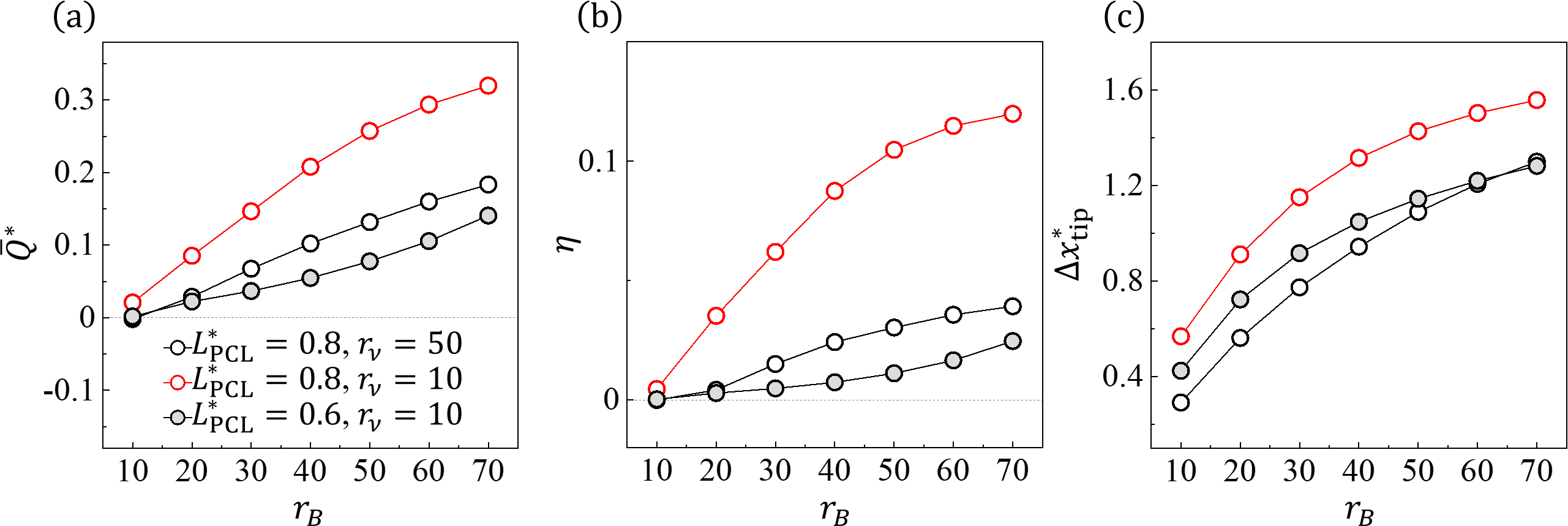}}
\caption{Variations of (a) $\bar{Q}^*$, (b) $\eta$, and (c) $\Delta x_{\mathrm{tip}}^*$ as a function of $r_B$ for different $L_\mathrm{PCL}^*$ and $r_{\nu}$.}
\label{fig_rb_q}
\end{figure}

The bending stiffness ratio ($r_B$) modulates the drag-elastic force balance and thereby affects both the filament beating pattern and fluid transport. To illustrate its impact, Fig. \ref{fig_rb_q} presents the variations of the time-averaged flow rate ($\bar{Q}^*$), transport efficiency ($\eta$), and tip amplitude ($\Delta x_{\mathrm{tip}}^*$) as functions of $r_B$. These quantities exhibit a similar trend, increasing with $r_B$ regardless of $L_\mathrm{PCL}^*$ and $r_{\nu}$. Notably, $\bar{Q}^*$, $\eta$, and $\Delta x_{\mathrm{tip}}^*$ attain substantially larger values for $L_\mathrm{PCL}^*=0.8$ and $r_{\nu}=10$, owing to the relatively low drag imposed by the ML. Moreover, at high $r_B$, the growth of $\Delta x_{\mathrm{tip}}^*$ slows down, as it becomes constrained by the filament length.

\begin{figure}
\centerline{\includegraphics[width=\linewidth]{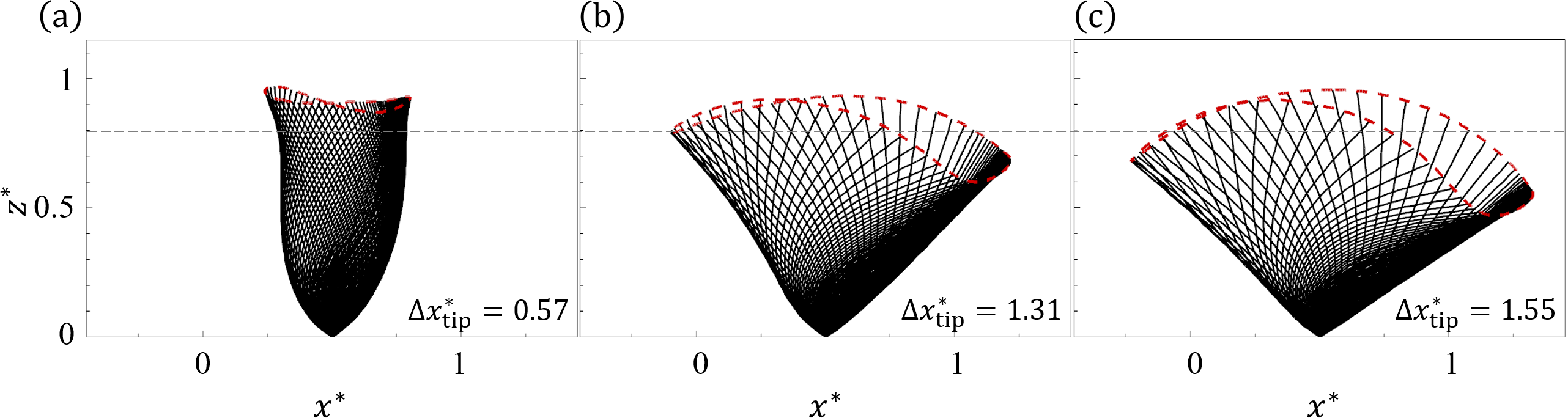}}
\caption{Superimposed instantaneous filament shapes in one beating period for (a) $r_B=10$, (b) $r_B=40$, and (c) $r_B=70$ ($L_\mathrm{PCL}^*=0.8$, $r_{\nu}=10$).}
\label{fig_rb_shape}
\end{figure}

\begin{table}[b]
\caption{Time-averaged flow rate ($\bar{Q}^*$), input power ($\bar{P}_{\mathrm{in}}^*$), effective kinetic energy of the fluid ($\bar{E}_{\mathrm{kf}}^*$), elastic strain energy of the filament ($\bar{E}_{\mathrm{es}}^*$), total fluid drag ($\bar{F}^{\mathrm{IB} *}_x$), ML-component of the fluid drag ($\bar{F}^{\mathrm{IB} *}_{x\mathrm{ML}}$) for different $r_B$ ($r_\nu = 10$, $L_\mathrm{PCL}^*=0.8$).}
\begin{ruledtabular}
\begin{tabular}{ccccccc}
\textrm{$r_B$}&
\textrm{$\bar{Q}^*$}&
\textrm{$\bar{P}_{\mathrm{in}}^*$} ($10^3$)&
\textrm{$\bar{E}_{\mathrm{kf}}^*$} ($10^3$)&
\textrm{$\bar{E}_{\mathrm{es}}^*$} ($10^3$)&
\textrm{$\bar{F}^{\mathrm{IB} *}_x$} ($10^3$)&
\textrm{$\bar{F}^{\mathrm{IB} *}_{x\mathrm{ML}}$} ($10^3$)\\
\colrule
10 & 0.021 & 1.24 & 0.006 & 0.25 & 1.58 & 0.55 \\
40 & 0.208 & 5.97 & 0.52 & 0.69 & 2.44 & 0.90 \\
70 & 0.320 & 10.27 & 1.23 & 0.85 & 2.87 & 1.04 \\
\end{tabular}
\end{ruledtabular}
\label{table_avevalue_rb}
\end{table}

\begin{figure}
\centerline{\includegraphics[width=\linewidth]{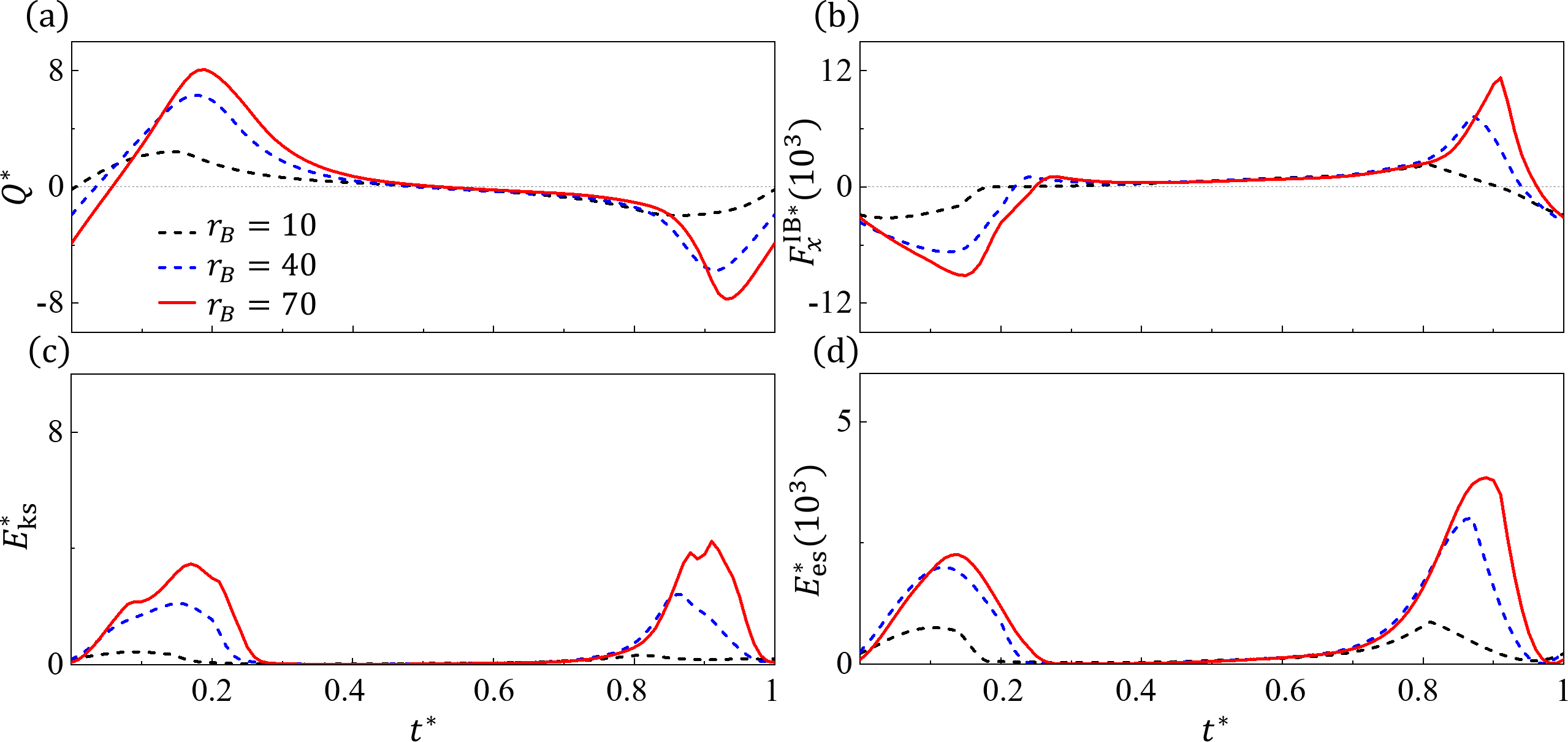}}
\caption{Time histories of (a) $Q^*$, (b) $F^{\mathrm{IB} *}_x$, (c) $E_{\mathrm{ks}}^*$, and (d) $E_{\mathrm{es}}^*$ for different $r_B$ ($L_\mathrm{PCL}^*=0.8$, $r_{\nu}=10$).}
\label{fig_rb_his}
\end{figure}

The rise in $\Delta x_{\mathrm{tip}}^*$ with increasing $r_B$ ($L_\mathrm{PCL}^*=0.8$, $r_{\nu}=10$) is visualized in Fig. \ref{fig_rb_shape}, accompanied by reduced deformation and enhanced asymmetry of the beating pattern. Physically, a higher $r_B$ strengthens the filament's resistance to fluid drag, allowing the drag-elastic force balance to be achieved with smaller deformation and thereby increasing the tip amplitude. Since $L_\mathrm{PCL}^*$ and $r_{\nu}$ are fixed, the effect of viscous diffusion remains unchanged with varying $r_B$, so that a larger tip amplitude consistently promotes stronger fluid propulsion. For $r_B=10$, the filament exhibits a nearly symmetric beating pattern and continuously penetrates the ML, yielding similar magnitudes of positive ($Q^{*}_{+}=0.988$) and negative ($Q^{*}_{-}=$ -0.979) flow rates, and thus a net flow rate close to zero. As $r_B$ increases, $Q^{*}_{+}$ rises while $Q^{*}_{-}$ decreases (Fig. \ref{fig_rb_his}(a)), with the increase in $Q^{*}_{+}$ outweighing the decrease in $Q^{*}_{-}$ due to the enhanced asymmetry, thereby raising the net flow rate. Filaments with higher $r_B$ also exhibit larger elastic strain energy and kinetic energy, which not only propel more fluid but also result in higher input power (Table \ref{table_avevalue_rb}; Fig. \ref{fig_rb_his}). Importantly, this additional input power is effectively converted into fluid kinetic energy, leading to enhanced net flow rate and transport efficiency (Fig. \ref{fig_rb_q}). This improvement can be attributed to the increased asymmetry of the beating pattern, which will be further discussed below.

\subsection{Relationships between flow rate and beating pattern}

\begin{figure}
\centerline{\includegraphics[width=\linewidth]{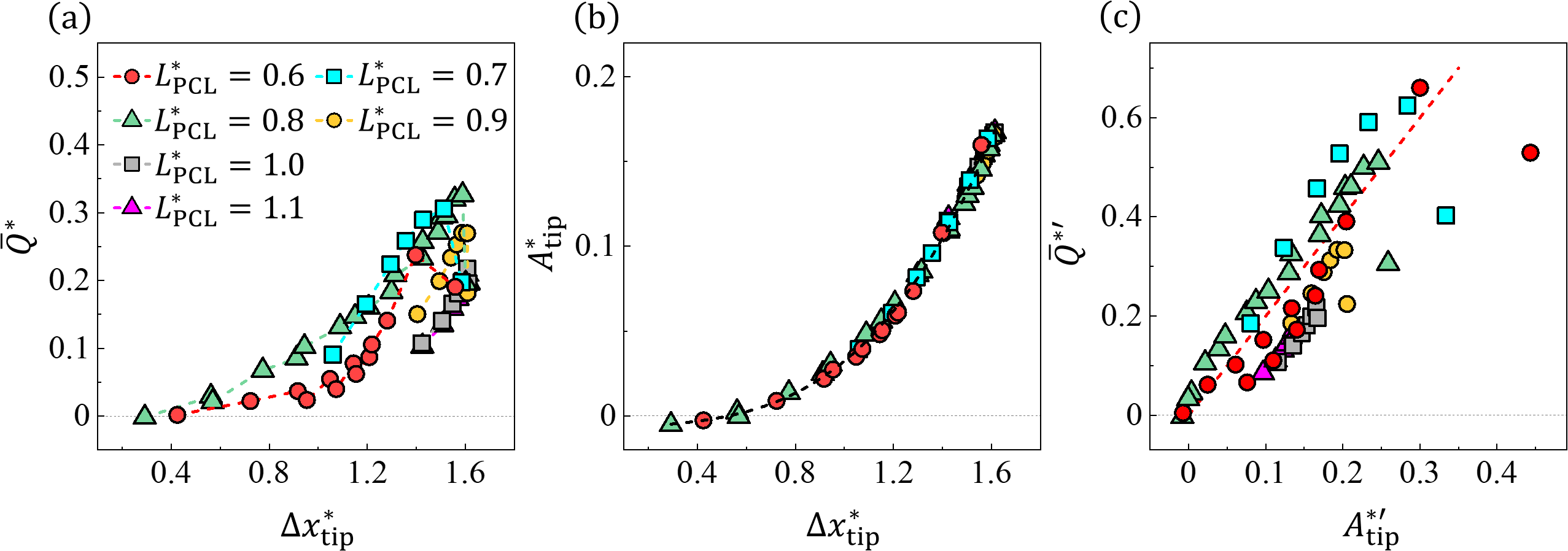}}
\caption{(a) Values of $\bar{Q}^*$ for various $\Delta x_{\mathrm{tip}}^*$. (b) Values of the area enclosed by the tip trajectory of the filament ($A_{\mathrm{tip}}^*$) for various $\Delta x_{\mathrm{tip}}^*$. The dashed line is obtained from a cubic polynomial fit. (c) Values of the re-normalized flow rate ($\bar{Q}^{* \prime}$) for various re-normalized area ($A_{\mathrm{tip}}^{* \prime}$).}
\label{fig_qvsx}
\end{figure}

According to the above discussions, the filament beating pattern plays a crucial role in determining the net flow rate of the ASL when varying $L_\mathrm{PCL}^*$, $r_{\nu}$, and $r_B$. The tip amplitude ($\Delta x_{\mathrm{tip}}^*$) serves as a key descriptor of the beating pattern, reflecting the drag-elastic force balance. Figure \ref{fig_qvsx}(a) presents the time-averaged flow rate ($\bar{Q}^*$) as a function of $\Delta x_{\mathrm{tip}}^*$ for all simulated cases. Data points corresponding to a given $L_\mathrm{PCL}^*$ are connected by dashed lines. For fixed $L_\mathrm{PCL}^*$, the net flow rate increases with $\Delta x_{\mathrm{tip}}^*$ up to a maximum and then decreases. The cases with $r_{\nu}=1$ exhibit the largest $\Delta x_{\mathrm{tip}}^*$ owing to the absence of the ML's retarding effect; however, the weakened viscous diffusion of momentum results in a low $\bar{Q}^*$. Similarly, cases with $L_\mathrm{PCL}^*=1.0$ and 1.1 also show large $\Delta x_{\mathrm{tip}}^*$, but their $\bar{Q}^*$ values are much smaller than those of other cases with comparable tip amplitudes.

Note that the tip trajectory of the filament is shown in each beating-pattern figure (e.g. Fig. \ref{fig_rb_shape}). To quantify the asymmetry of the beating, we define the normalized area enclosed by the tip trajectory as $A_{\mathrm{tip}}^*=A_{\mathrm{tip}}/L^2$. The value of $A_{\mathrm{tip}}^*$ is negative when the trajectory of the recovery stroke lies above that of the power stroke, reflecting the difference between the regions swept during the two strokes. A larger $A_{\mathrm{tip}}^*$ corresponds to stronger forward propulsion of the fluid. Figure \ref{fig_qvsx}(b) plots $A_{\mathrm{tip}}^*$ against $\Delta x_{\mathrm{tip}}^*$. The dashed line, obtained from a cubic polynomial fit, shows that $A_{\mathrm{tip}}^*$ increases with $\Delta x_{\mathrm{tip}}^*$, indicating enhanced asymmetry.

Considering that $L_\mathrm{PCL}^*$ strongly influences both the flow rate and the beating pattern, we adopt $L_\mathrm{PCL}$ as the reference length instead of $L$ to re-normalize $\bar{Q}^*$ and $A_{\mathrm{tip}}^*$. After re-normalization, the values of $\bar{Q}^{* \prime}$ as functions of $A_{\mathrm{tip}}^{* \prime}$ are plotted in Fig. \ref{fig_qvsx}(c). The results reveal a nearly linear dependence of the flow rate on the asymmetry when the ML viscosity exceeds that of the PCL. This demonstrates that variations in beating asymmetry act as a direct geometric determinant of both net flow rate and transport efficiency.

\section{Conclusions}\label{sec:conclusions}

In this study, we conducted numerical simulations to investigate fluid transport driven by an active filament in a three-dimensional two-phase flow. The Shan-Chen multiphase model was incorporated into a coupled immersed boundary-lattice Boltzmann solver, enabling two-way coupling between the flexible filament and the surrounding fluid. Parametric studies were carried out over a wide range of periciliary layer (PCL) thicknesses ($L^*_{\mathrm{PCL}}$), viscosity ratios ($r_{\nu}$) between the PCL and mucus layer (ML), and filament bending stiffness ratios ($r_B$). The influences of these parameters on both filament dynamics and fluid transport were systematically examined.

The filament propels the fluid forward during the power stroke and backward during the recovery stroke, yielding a positive time-averaged flow rate ($\bar Q^*$) as a consequence of its spatially asymmetric beating. Within the parameter ranges investigated, a moderate $L^*_{\mathrm{PCL}}$ and $r_{\nu}$ combined with a high $r_B$ tend to yield higher net flow rate and transport efficiency ($\eta$). This arises from the interplay of two competing mechanisms: the drag-elastic force balance and the viscous diffusion of momentum. The former controls the tip amplitude and filament deformation, thereby determining the ability to generate propulsion, while the latter governs the extent to which the flow responds to filament motion. Better fluid transport tends to occur when the filament slightly penetrates into the ML with moderate viscosity, allowing relatively low drag while maintaining effective filament-ML interaction. Under this condition, the input power is more efficiently converted into fluid kinetic energy, leading to a high $\bar Q^*$ and $\eta$. Furthermore, $\bar Q^*$, $\eta$ and $\Delta x^*_{\mathrm{tip}}$ each increase monotonically with increasing $r_B$, irrespective of $L^*_{\mathrm{PCL}}$ and $r_{\nu}$. Since viscous diffusion remains unchanged when $L^*_{\mathrm{PCL}}$ and $r_{\nu}$ are fixed, a larger tip amplitude consistently enhances fluid propulsion. The relationship between flow rate and beating pattern was also clarified: $\bar Q^*$ increases and then decreases with increasing $\Delta x^*_{\mathrm{tip}}$. To characterize asymmetry, the area enclosed by the tip trajectory ($A_{\mathrm{tip}}^*$) was introduced, which increases with $\Delta x^*_{\mathrm{tip}}$. A larger $A_{\mathrm{tip}}^*$ corresponds to stronger forward propulsion. Finally, the results reveal a nearly linear dependence of the re-normalized flow rate ($\bar{Q}^{* \prime}$) on the re-normalized trajectory area ($A_{\mathrm{tip}}^{* \prime}$) when the ML viscosity exceeds that of the PCL, demonstrating that variations in beating asymmetry act as a direct geometric determinant of both net flow rate and transport efficiency.

The present study employs a simplified representation of mucociliary clearance. Future work will incorporate the non-Newtonian rheology of mucus, ciliary coordination, and fluid feedback on beating frequency, with the aim of considering more realistic and physiologically relevant conditions for mucociliary clearance.

\begin{acknowledgments}
Centre de Calcul Intensif d'Aix-Marseille University is acknowledged for granting access to its high performance computing resources. The authors thank Dr. G. Wang for his kind assistance in the code validation. This work was supported by the BronchoClogDrain project (ANR-22-CE30-0045) funded by the French National Research Agency (ANR).
\end{acknowledgments}

\bibliography{apssamp}

\end{document}